\documentclass[twocolumn]{aastex63}

\usepackage{savesym}
\savesymbol{tablenum}
\usepackage{siunitx}
\restoresymbol{SIX}{tablenum}
\usepackage[caption=false]{subfig}
\usepackage{amsmath}
\usepackage{epstopdf}
\usepackage[utf8]{inputenc}

\submitjournal{ApJ}
\accepted{2020/08/19}

\shorttitle{The Equatorial Jet Speed on Tidally Locked Planets: I -- Terrestrial Planets}
\shortauthors{Hammond, Tsai, \& Pierrehumbert}

\begin{document}

\title{The Equatorial Jet Speed on Tidally Locked Planets: I -- Terrestrial Planets}
\correspondingauthor{Mark Hammond}
\email{markhammond@uchicago.edu}

\author[0000-0002-6893-522X]{Mark Hammond}
\affiliation{Department of the Geophysical Sciences \\
University of Chicago \\
5734 South Ellis Avenue, Chicago, IL 60637, USA}
\affiliation{Department of Physics \\
University of Oxford \\
Parks Road, Oxford, UK}

\author[0000-0002-8163-4608]{Shang-Min Tsai}
\affiliation{Department of Physics \\
University of Oxford \\
Parks Road, Oxford, UK}

\author[0000-0002-5887-1197]{Raymond T. Pierrehumbert}
\affiliation{Department of Physics \\
University of Oxford \\
Parks Road, Oxford, UK}


\begin{abstract}

The atmospheric circulation of tidally locked planets is dominated by a superrotating eastward equatorial jet. We develop a predictive theory for the formation of this jet, proposing a mechanism in which the three-dimensional stationary waves induced by the day-night forcing gradient produce an equatorial acceleration. This is balanced in equilibrium by an interaction between the resulting jet and the vertical motion of the atmosphere. The three-dimensional structure of the zonal acceleration is vital to this mechanism.

We demonstrate this mechanism in a hierarchy of models. We calculate the three-dimensional stationary waves induced by the forcing on these planets, and show the vertical structure of the zonal acceleration produced by these waves, which we use to suggest a mechanism for how the jet forms. GCM simulations are used to confirm the equilibrium state predicted by this mechanism, where the acceleration from these waves is balanced by an interaction between the zonal-mean vertical velocity and the jet. We derive a simple model of this using the ``Weak Temperature Gradient'' approximation, which gives an estimate of the jet speed on a terrestrial tidally locked planet.

We conclude that the proposed mechanism is a good description of the formation of an equatorial jet on a terrestrial tidally locked planet, and should be useful for interpreting observations and simulations of these planets. The mechanism requires assumptions such as a large equatorial Rossby radius and weak acceleration due to transient waves, and a different mechanism may produce the equatorial jets on gaseous tidally locked planets.

\end{abstract}


\section{Introduction} \label{sec:intro}

Tidally locked planets always present the same face to the star that they orbit. Their atmospheric circulation is dominated by an equatorial jet, the strength of which determines directly observable features like the hot-spot shift and day-night contrast. \citet{showman2011superrotation} showed that the jet is produced by the day-night instellation gradient, which induces stationary equatorial waves that transport prograde momentum towards the equator. 

No studies so far have used this process to predict the equilibrium jet speed on these planets, as the process that balances this acceleration has not been identified. In this study we propose a mechanism by which this jet forms on terrestrial tidally locked planets, which does not rely on frictional drag. This provides a estimate of the jet speed that only depends on the basic atmospheric and planetary parameters. Our primary aim is to demonstrate the mechanism by which the jet forms, and to derive how the jet speed with the planetary parameters. Our estimate of an exact jet speed only applies to the idealised atmospheres we consider in this study, and will not apply to planets with thick atmospheres, significantly different heating profiles, or strong moisture effects.

The structure of this paper is as follows. In Section \ref{sec:global-circulation-tl-planets} we review previous work on the atmospheric circulation of tidally locked planets, and show their typical global circulation in a GCM simulation. In Section \ref{sec:3d-stationary-wave-response} we introduce an idealised model of the three-dimensional stationary waves induced in the atmosphere of a tidally locked planets by its day-night instellation gradient. The mechanism we propose for the formation of the jet relies entirely on the zonal acceleration caused by these stationary waves, so we aim to isolate them in this idealised model. We solve the primitive equations on a beta-plane using the Dedalus software package, and show the structure of these waves and the zonal acceleration that they produce. The vertical profile of the zonal acceleration is then used in Section \ref{sec:jet-formation-mechanism} to propose a mechanism for the formation and equilibration of the jet. \citet{leovy1987zonal} and \citet{zhu2006maintenance} proposed similar mechanisms for the formations of zonal jets on Venus and Titan.

Section \ref{sec:gcm-simulation-results} uses a suite of GCM simulations of terrestrial tidally locked planets to test the proposed theory. We show that the equilibrium zonal momentum budget matches the expected balance from the proposed mechanism. We also show that the scaling of equatorial jet speed with instellation approximately matches the predicted speed from the stationary wave calculation. In Section \ref{sec:predicting-jet-speed} we derive a simple estimate of the maximum equatorial jet speed on a terrestrial tidally locked planet:

\begin{equation}
    \overline{u} \sim \frac{2.53 \pi ag^{3}\sigma^{1/2}}{RN_{*}^{2}p_{0}c_{p}}  F_{0}^{1/2},
\end{equation}

for planetary radius $a$, acceleration due to gravity $g$, Stefan-Boltzmann constant $\sigma$, specific gas constant $R$, Brunt–Väisälä frequency $N_{*}$, surface pressure $p_{0}$, specific heat capacity $c_{p}$, and instellation $F_{0}$. This estimate corresponds to the jet speed at a height $z=H$, where $H$ is the atmospheric scale height. The constant of proportionality depends on the heating profile at the substellar point, so the predicted jet speed is a scaling relation in general, and only a numerical prediction for planets with zero albedo and a forcing profile with a vertical wavelength $\approx 2 H$.

In Section \ref{sec:discussion} we discuss how this mechanism relies on several assumptions and simplifications, and suggest how other sources of acceleration such as transient waves could affect the jet speed. We also show how the different properties of gaseous ``hot Jupiter'' exoplanets could complicate the formation of a jet via this mechanism, which we will investigate in a forthcoming study.

We conclude that this mechanism is a good description of the formation of the equatorial jet on a terrestrial tidally locked planet with a dry, cloud-free atmosphere, and can predict the approximate jet speed for these planets. The key assumptions required by this mechanism are that the zonal acceleration is initially dominated by the contribution from stationary waves, and that once the jet forms it does not strongly affect the magnitude of this zonal acceleration. This mechanism could be extended to describe the jet formation and speed on planets with thicker atmospheres, clouds with strong radiative effects, or significant moisture content.

\section{The Global Circulation of Terrestrial Tidally Locked Planets}\label{sec:global-circulation-tl-planets}

Many of the best planetary candidates for atmospheric characterisation or potential habitability are expected to be tidally locked, but their atmospheric circulation is not fully understood. This section reviews previous work related to the equatorial jets of terrestrial tidally locked planets, and shows the typical features of a simulation of the global circulation on such a planet.

\subsection{Review of Previous Work}

The atmospheric circulation of tidally locked planets is measurable through observations such as thermal phase curves \citep{parmentier2017handbook}. This study is motivated by the need to understand the formation of their equatorial jets, which are the dominant dynamical feature of this circulation \citep{pierrehumbert2018review}. Numerical atmospheric modelling has been invaluable to understanding the composition and climates of tidally locked gaseous planets such as hot Jupiters \citep{mayne2014unified,showman2015circulation,drummond2018observable,mendoncca2018revisiting,debras2020acceleration} and terrestrial planets \citep{joshi1997tidally,hammond2017climate,boutle2017proxima}. The lack of observational data makes verifying the accuracy of these models difficult, compared to models of the Earth and the other planets of the Solar System. This has led to studies using different modelling approaches \citep{cho2015sensitivity} and different approximations \citep{mayne2019limits} as it is not clear what techniques give the most realistic results. This study aims to provide a theoretical basis for the equatorial jet on these planets, which may help to guide modelling choices -- for example, ensuring that an imposed surface drag or top-of-atmosphere sponge layer does not interfere with the momentum balance of the jet, or using a high enough upper boundary to fully resolve the zonal momentum fluxes that produce the jet.
 
Previous studies have demonstrated different circulation regimes on these planets, varying properties such as instellation and rotation rate to show their effect on the global circulation and its observable features \citep{kataria2014atmospheric, showman2015circulation, carone2015regimes}. Other studies have compared suites of simulations to observations such as phase curves of terrestrial planets \citep{demory201655cnce, hammond2017climate} and hot Jupiters \citep{arcangeli2019climate}. More detailed measurements of circulation are becoming possible, such as measuring wind speeds via Doppler spectroscopy \citep{louden2015spatially, brogi2016rotation,flowers2019high}, and measuring multi-dimensional temperature maps by eclipse mapping and spectral phase curves \citep{majeau20122dmap,stevenson2014thermal}. 

An eastward superrotating equatorial jet is a common feature of almost all simulations of tidally locked planets. \citet{read2018superrotation} define local superrotation as a state with a local excess of eastward atmospheric angular momentum relative to solid-body rotation at the equator. Any eastward flow at the equator is therefore superrotating and must be produced by up-gradient angular momentum transport towards the equator, a requirement know as Hide's Theorem \citep{hide1969dynamics}. This process has been investigated for Solar System planets such as Venus (e.g. \citep{fels1974interaction}) and the Earth (e.g. \citep{shell2004superrotation}). \citet{showman2010superrotation} showed that on tidally locked planets this transport is provided by planetary-scale stationary waves, similar to the equatorial waves present in the tropics of the Earth \citep{matsuno1966quasi}. In a pioneering study, \citet{showman2011superrotation} developed this concept further and demonstrated it in a 2D linear shallow-water model, a 2D nonlinear shallow-water model and a 3D GCM. Their linear model relied on a linear drag to produce the appropriate stationary waves for an eastward equatorial acceleration (specifically, the equatorial Kelvin wave), and did not produce a zonal equatorial acceleration without this linear drag. Their non-linear model did produce a zonal equatorial acceleration without linear drag, which we will explore further in Section \ref{sec:stationary-wave-structure}. \citet{heng2014analytical} and \citet{perez2013atmospheric} further explored two-dimensional models of this system, showing how non-linear balance or a more realistic forcing field gives different solutions, while still preserving the formation of the equatorial jet. This study focuses on the three-dimensional structure of the zonal acceleration, following authors such as \citet{mendoncca2020angular} and \citet{debras2020acceleration} who analysed the vertical structure of the zonal acceleration on tidally locked planets, identifying the key role of the vertical transport of zonal momentum.

\citet{tsai2014three} followed the approach of \citet{wu2000vertical} to construct a three-dimensional linear model where the stationary wave response to stellar forcing is composed of separable vertical modes coupled to two-dimensional shallow-water systems. This showed that a uniform eastward zonal flow shifts the stationary equatorial waves eastward, producing the equatorial hot-spot shift seen in observations \citep{parmentier2017handbook}. \citet{tsai2014three} also showed that on hot Jupiters the eastward shift of these equatorial waves can reduce the acceleration they produce, allowing the equatorial jet to reach a steady state. \citet{hammond2018wavemean} used a two-dimensional linear shallow-water model to show how a zonal flow with meridional shear, and an associated geostrophically balanced geopotential perturbation, produces the characteristic shape of the atmospheric circulation and hot-spot shift on tidally locked planets.

The global circulation on a slowly rotating tidally locked planet has similarities to the tropical circulation on the Earth due to the large Rossby number in both cases \citep{pierrehumbert2018review}. This leads to behaviour that can be approximated by the ``Weak Temperature Gradient'' (WTG) regime \citep{pierrehumbert2010palette,koll2016temperature}, where a non-linear balance in the zonal momentum equation leads to weak horizontal geopotential gradients (and therefore weak temperature gradients). This will be a key simplifying assumption later in this study. The tropics of the Earth also host similar equatorial stationary waves to those on tidally locked planets \citep{matsuno1966quasi,gill1980solutions}, which can lead to similar equatorial superrotation \citep{norton2006tropical}. \citet{lutsko2018response} considered a very similar system to this study, with a localised tropical heat source on the equator of the Earth rather than a planetary-scale day-night instellation gradient.

This study also builds on other work on the magnitude and scaling behaviour of the velocity and temperature fields on these planets. \citet{komacek2016daynighti} and \citet{komacek2017daynightii} introduced a predictive theory for the scaling of temperature and velocity perturbations in the atmospheres of tidally locked planets, and successfully applied it to explain the observed scaling of day-night temperature differences on hot Jupiters. \citet{koll2015phasecurves} produced a similar theory for terrestrial planets based on the WTG approximation, which \citet{kreidberg2019lhs} used to interpret observations of the thermal phase curve of a terrestrial planet. \citet{zhang2017dynamics} derived scaling relations for properties of the atmospheric circulation of tidally locked planets, based on the relations of \citet{komacek2016daynighti}. \citet{koll2018heatengine} treated the global circulation of a tidally locked planet as a heat engine, to predict the wind speeds on hot Jupiters. Many of these theoretical predictions of observable quantities depend on the equatorial jet speed, which previously needed to be diagnosed from GCM simulations. This study aims to provide a predictive theory for this jet speed on terrestrial tidally locked planets, to enable the prediction of many other observable quantities.

\begin{figure*}
\centering 
\subfloat[Geopotential and velocity fields at the \SI{390}{\milli\bar} level]{%
  \includegraphics[width=\columnwidth]{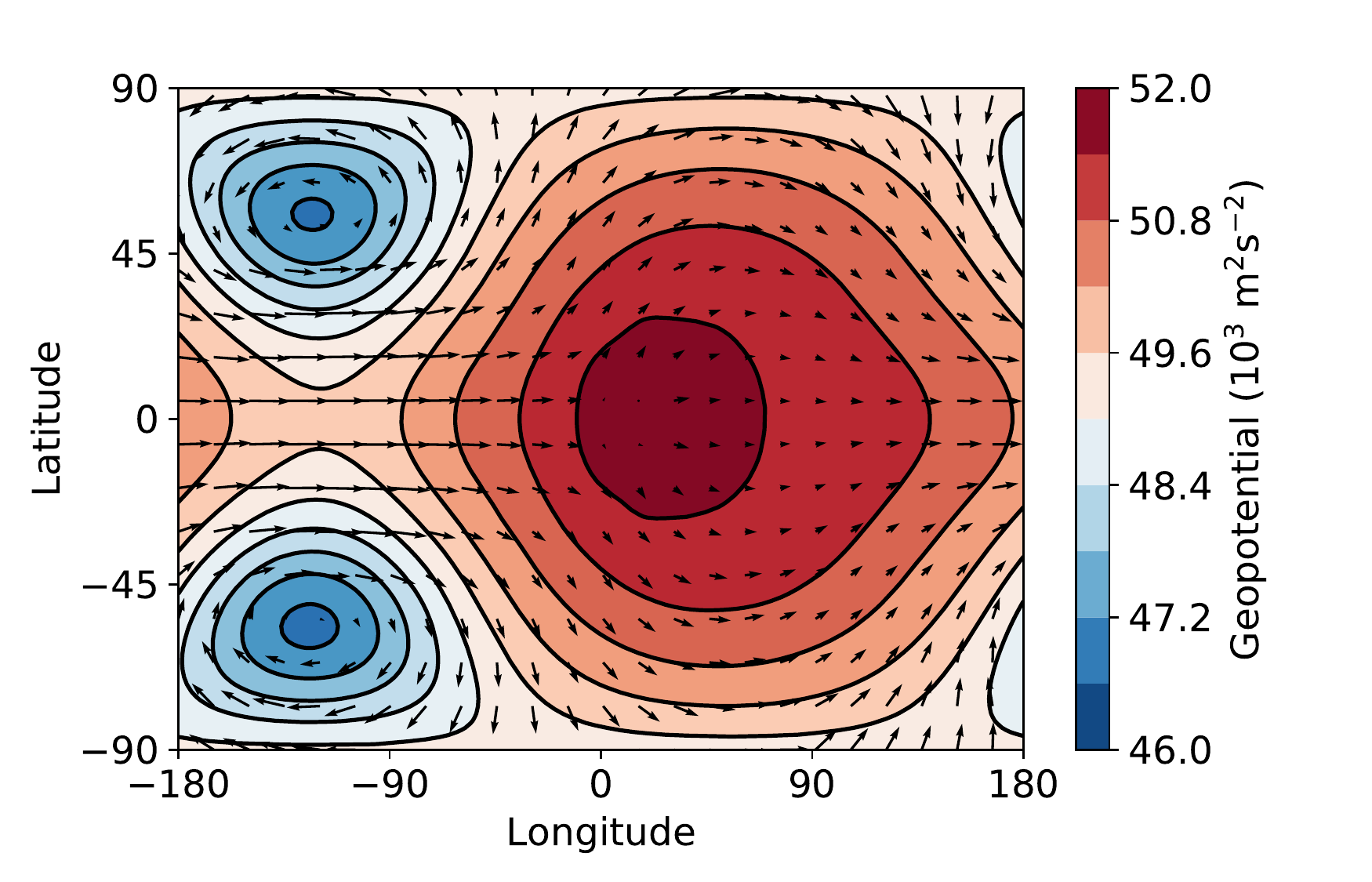}%
  \label{fig:gcm-control-temperature-quivers}%
}
\subfloat[Zonal-mean zonal velocity]{%
  \includegraphics[width=\columnwidth]{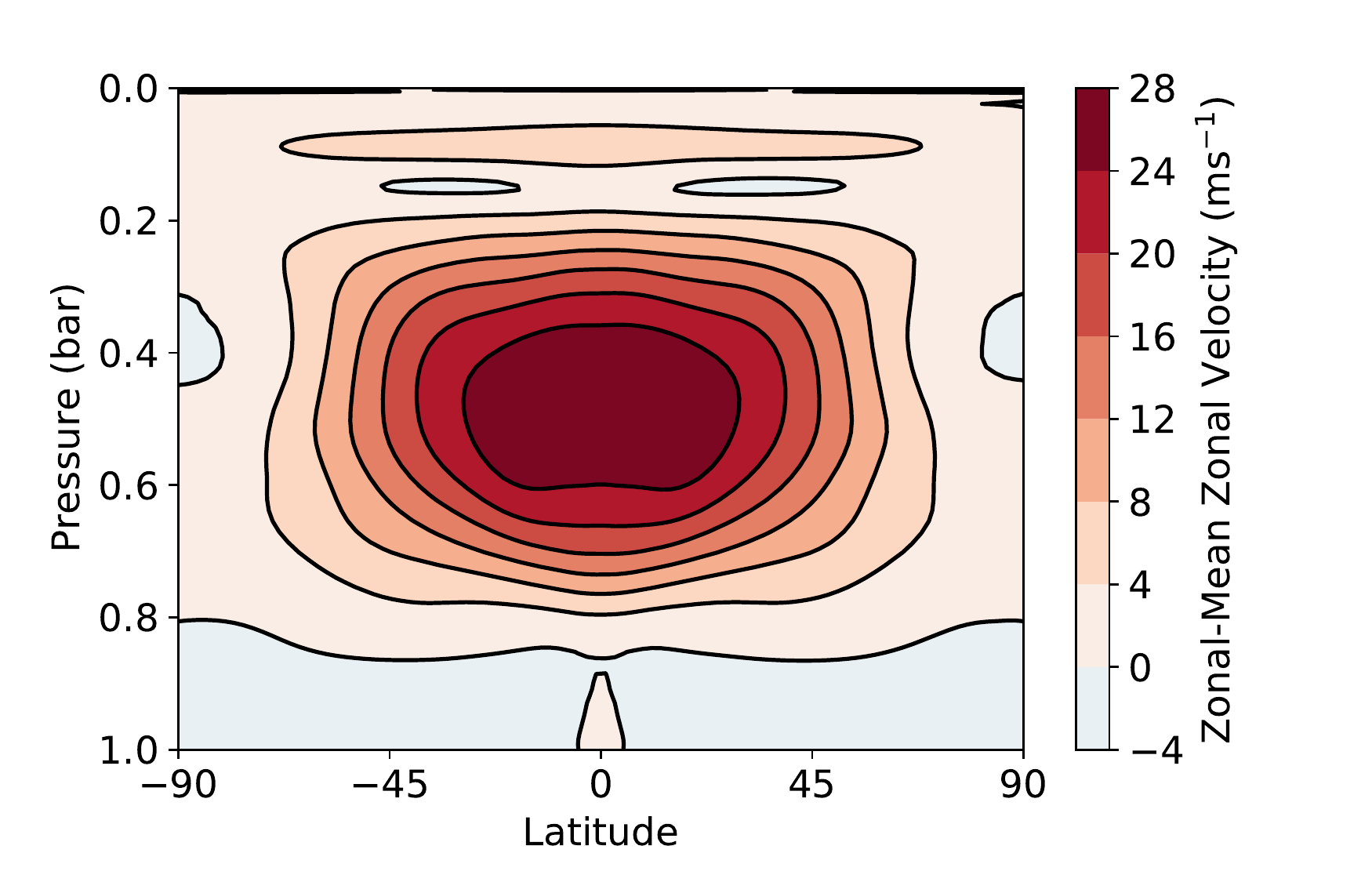}%
  \label{fig:gcm-control-zonal-flow}%
}
\caption{The equilibrium circulation of a simulation of a terrestrial tidally locked planet with instellation \SI{300}{\watt\per\metre\squared}, discussed in Section \ref{subsec:example-circulation}. It is plotted at the \SI{390}{\milli\bar} level, for consistency with later figures. The substellar point is located at \ang{0} longitude. This is a typical circulation pattern for a tidally locked planet, with an eastward equatorial jet producing an eastward hot-spot shift. The eastward equatorial jet is centered at about \SI{500}{\milli\bar}.}\label{fig:gcm-control-circulation}
\end{figure*}

\subsection{Typical Global Circulation}\label{subsec:example-circulation}

Figure \ref{fig:gcm-control-circulation} shows the typical global circulation on a terrestrial tidally locked planet, similar to the planets of the Trappist-1 system \citep{gillon2017seven}. The simulation was run in the GCM Exo-FMS \citep{ding2016convection,pierrehumbert2016dynamics,hammond2017climate,hammond2018wavemean,pierrehumbert2018review}. Section \ref{sec:numerical-simulations} describes the details of the numerical modelling in this study in more depth; we introduce a single simulation here to show the key features of its circulation. Its main parameters are a radius of $6\times10^{6}$ km, a rotation period of 10 days, and instellation at the substellar point of $300$ Wm$^{-2}$. 

Figure \ref{fig:gcm-control-temperature-quivers} shows the geopotential and velocity fields at the \SI{500}{\milli\bar} pressure level of this simulation, at the peak of the equatorial jet. \citet{tsai2014three} and \citet{hammond2018wavemean} explained that the geopotential and velocity fields represent a ``Matsuno-Gill'' pattern of stationary equatorial waves \citep{matsuno1966quasi,gill1980solutions}, which are shifted eastwards by the equatorial jet. \citet{hammond2018wavemean} showed how the velocity field is primarily a combination of a meridionally sheared (but zonally uniform) equatorial jet, plus a stationary wave response with zonal wavenumber 1. This results in a strong eastward velocity at \ang{-90} where these two components combine, and a region of weak flow at \ang{90} where these two components cancel. 

The eastward shift of the peak of the geopotential in Figure \ref{fig:gcm-control-temperature-quivers} corresponds to a shift in the peak of the temperature field eastwards from the substellar point. The maximum shift in the geopotential is at the level of the peak of the equatorial jet; the maximum shift in the temperature field is at a different pressure level as it is out of vertical phase with the geopotential field due to the hydrostatic relation. The shift of the hot-spot and the difference in temperature between the day-side and the night-side are observable quantities \citep{parmentier2017handbook,komacek2016daynighti}, which should depend on the speed of this jet \citep{zhang2017dynamics}.  The next section introduces the idealised model that we use to calculate the three-dimensional stationary wave response to a day-night instellation gradient.

\section{Three-Dimensional Stationary Wave Response to Forcing}\label{sec:3d-stationary-wave-response}

In this section we calculate the three-dimensional stationary waves produced in the primitive equations on an equatorial beta-plane by an idealised forcing, representing the atmosphere of a tidally locked planet. Our aim is to show the structure of the zonal acceleration produced by these stationary waves, which we will use in the next section to suggest a mechanism for the formation and equilibration of the equatorial jet. Unlike previous studies, we find the response to forcing without imposing a linear drag on the horizontal velocities, which we will show to be vital to matching the magnitude of the velocity perturbations and zonal acceleration in our GCM simulations.

\subsection{Idealised Beta-Plane Model}\label{sec:idealised-beta-plane-model}

The adiabatic, inviscid primitive equations in height (log-pressure) coordinates $(x,y,z)$ are \citep{vallis2006book}:

\begin{equation}
    \begin{aligned}
        \frac{\mathrm{D} \boldsymbol{u}}{\mathrm{D} t}+f \times \boldsymbol{u}&=-\nabla_{z} \Phi, \\
        \frac{\partial \Phi}{\partial z}&=\frac{R T}{H}, \\
        \nabla_{z} \cdot \boldsymbol{u}+\frac{1}{\rho} \frac{\partial (\rho w)}{\partial z}&=0, \\
        \frac{\partial T}{\partial t}+\boldsymbol{u} \cdot \nabla_{z} T +\frac{H}{R} N_{*}^{2}w &=0,
    \end{aligned}
\end{equation}

where the advective derivative is:

\begin{equation}
    \frac{\mathrm{D}}{\mathrm{D} t}=\frac{\partial}{\partial t}+\boldsymbol{u} \cdot \nabla_{z}+ w \frac{\partial}{\partial z},
\end{equation}

and the horizontal derivative at constant $z$ is:

\begin{equation}
    \nabla_{z} = (\frac{\partial}{\partial x},\frac{\partial}{\partial y}).
\end{equation}

The variables in these equations are the temperature $T$, the horizontal velocity $\boldsymbol{u}=(u,v)$, the vertical velocity $w$, and the geopotential $\Phi$. The log-pressure $z$ coordinate can be transformed to the pressure coordinate $p=p_{s}e^{-z/H}$, and the vertical velocity $w$ can be transformed to the vertical pressure velocity $\omega = -(p/H)w$.

The Coriolis parameter is $f = 2 \Omega \sin{\phi}$, where $\Omega$ is the planetary rotation rate and $\phi$ is the latitude (which we will represent with the $y$ coordinate on the beta-plane, so that $f = 2 \Omega \sin{\phi} = \beta y$). The density is $\rho=\rho_{0} \exp (-z / H)$, for a surface density $\rho_{0}$ (determined from the surface pressure, for an ideal gas) and a scale height that we approximate as $H = RT_{0}/g$. $R$ is the specific gas constant, $g$ is the acceleration due to gravity, and $T_{0}$ is the planetary equilibrium temperature for the instellation $F_{0}$. The Brunt–Väisälä frequency $N_{*}$ is defined by $N_{*}^{2} = \frac{R}{H}(\kappa T / H+\partial T / \partial Z)$, where the dry adiabatic exponent $\kappa = R/c_{p}$ for the specific heat capacity $c_{p}$. We approximate $N_{*}^{2}$ to have constant value of \SI{5e-4}{\per\second\squared}, to approximate the value used in the tropics of the Earth by \citet{wu2000vertical}, and to match the magnitude of the velocity perturbations in our GCM simulations in Section \ref{sec:gcm-simulation-results}.

We recast these equations onto the beta-plane, add a forcing $Q$ to represent that on a tidally locked planet, and introduce a linear radiative cooling term and a Rayleigh drag term (which we set to zero for the ``non-linear'' solutions later). We impose a second-order divergence damping, second-order hyperdiffusion, and a sponge layer to stabilise the horizontal velocity fields \citep{jablonowski2011pros}, which are all represented by the terms $D_{\boldsymbol{u}}$ and $D_{T}$. This results in the system:

\begin{equation}\label{eqn:dedalus-equations}
    \begin{aligned}
\frac{\mathrm{D} \boldsymbol{u}}{\mathrm{D} t} + \alpha_{dyn}\boldsymbol{u} + \beta \boldsymbol{z} \times \boldsymbol{u} &=-\nabla_{z} \Phi + D_{\boldsymbol{u}}, \\ 
        \frac{\partial \Phi}{\partial z} &= \frac{R T}{H}, \\
        \nabla_{z} \cdot \boldsymbol{u}+\frac{\partial w}{\partial z}-\frac{w}{H} &= 0, \\
        \frac{\partial T}{\partial t}+\alpha_{rad}\boldsymbol{T} +\boldsymbol{u} \cdot \nabla_{z} T +\frac{H}{R} N_{*}^{2} w&= Q + D_{T},
    \end{aligned}
\end{equation}

where $\alpha_{rad} = 1/\tau_{rad}$, for a constant radiative timescale $\tau_{rad}$:

\begin{equation}
    \tau_{rad} = \frac{e^{-2/3} p_{s} c_{p}}{4 g \sigma T_{0}^{3}},
\end{equation}

for an equilibrium temperature $T_{0}$ defined by $F_{0} = \sigma T_{0}^{4}$. The factor $ e^{-2/3} p_{s}$ corresponds to the pressure level at which the longwave optical depth is $2/3$ in our GCM simulations, i.e. the radiating level of the outgoing longwave radiation (the structure of the solutions is not directly sensitive to this parameter). The dynamical damping rate $\alpha_{dyn}$ is an arbitrary parameter corresponding to the rate of a linear Rayleigh drag, which we set to zero for most of the calculations.

The second-order diffusive damping and second-order divergence damping applied to the velocity field has the form:

\begin{equation}
    D_{\boldsymbol{u}} = S(z) K_{2} \nabla(\boldsymbol{u}) + \nu_{2}\Delta(\Delta\cdot\boldsymbol{u}),
\end{equation}

and the second-order diffusive  damping applied to the temperature field is:

\begin{equation}
    D_{T} = S(z) K_{2} \nabla(T),
\end{equation}

where the vertical distribution of the second-order diffusion terms is:

\begin{equation}
    S(z) = 1+5\left(1+\tanh\left(\frac{z-z_{s}}{z_{w}}\right)\right),
\end{equation}

to give a sponge layer that prevents the reflection of waves in the vertical direction from the upper boundary of the model. The sponge layer height is $z_{s} = 3H$ and the length scale over which the sponge layer increases to its maximum value is $z_{w} = H/3$. We choose the coefficients to be $K_{2} = 10^{7}$ and $\nu_{2} = 3 \times 10^{7}$ to stabilise the calculations, without playing a major role in the momentum balance \citep{jablonowski2011pros}.

The forcing $Q$ is a three-dimensional field representing the heating applied to the atmosphere on a terrestrial tidally locked planet. We use an idealised vertical heating profile similar to that used by \citet{wu2000vertical} to represent heating by convective plumes in the tropics of the Earth, as the atmosphere in the simulations in this study is forced by absorption of longwave radiation from the surface and by dry convective adjustment:

\begin{equation}
    Q_{z}(z)=\begin{cases}
        \sin( \frac{\pi z}{H} ) &z < H\\
        0 &z > H\\
    \end{cases}
\end{equation}

\citet{gill1986nonlinear} uses the same profile to represent localised tropical heating on the Earth that excites waves with a vertical wavelength $2H$. We found that the qualitative structure of the resulting stationary waves did not depend strongly on the exact structure of the vertical heating profile, as long as the forcing primarily excited waves with this vertical wavelength. The discontinuity in the first derivative of this vertical profile did excite high-order modes than smoother forcing profiles; we still used this profile as we found that the abrupt end to the forcing at a height $H$ produced zonal acceleration profiles that matched those in the later GCM simulations well. In \citet{tsai2014three} a similar system of stationary waves is forced by a well-defined heating profile due to absorption of stellar radiation in the atmosphere of a hot Jupiter.

The forcing has the same horizontal distribution at all vertical levels, following the cosine of the longitude $\theta$ to represent the instellation on a tidally locked planet, and following a Gaussian envelope in the meridional direction:

\begin{equation}
    Q_{xy}(x,y) = \cos(\theta) e^{-\frac{y^{2}}{y_{0}^{2}}},
\end{equation}

where the latitudinal scale of the forcing $y_{0}$ is set to be $\sqrt{2}a$, where $a$ is the radius of the planet, in order to generate planetary-scale stationary waves similar to those in the GCM simulations and the shallow-water model of \citet{showman2011superrotation}.

We normalise the three-dimensional forcing field $Q = Q_{x,y,z} = Q_{0}Q_{z}(z)Q_{xy}(x,y)$ such that the column-integrated energy absorbed at the substellar point is equal to the instellation at the top of the atmosphere; i.e. that there is zero albedo and all of the instellation acts to heat the atmosphere. For an instellation $F_{0}$, this requires:
\begin{equation}
    \begin{aligned}
        F_{0} &= \int Q_{0} Q_{z}(z) c_{p} \rho(z) d z \\
        &=\int_{0}^{H} Q_{0} \sin \left(\frac{\pi z}{H}\right) c_{p} \rho_{0} e^{-z/H} d z \\
        &= \frac{(1+e)\pi}{e(1+\pi^{2})}\ c_{p} \rho_{0} H Q_{0} \\
        &\approx \frac{0.395 c_{p} p_{0}}{g} Q_{0}.
    \end{aligned}  
\end{equation}

For a surface density $\rho_{0} = p_{s}M/(RT_{0})$ and a scale height $H=RT/Mg$, this gives $Q_{0} \approx \frac{2.53 g F_{0}}{p_{s}c_{p}}$. The constant of proportionality depends on the vertical profile of heating at the substellar point, and the albedo of the planet, so in general this is a scaling relation $Q_{0} \sim \frac{g F_{0}}{p_{s}c_{p}}$ rather than an exact equality.

We solve Equation \ref{eqn:dedalus-equations} in a three-dimensional spectral basis using the Dedalus package \citep{burns2016dedalus,burns2020dedalus}. This is an open-source Python package for the numerical solution of partial differential equations with spectral methods. We find the stationary solution of this forced system using different basis functions in each of its three dimensions. The $x$ direction is represented with a Fourier basis, from which we exclude all zonally uniform modes with zonal wavenumber $n=0$. This prevents the formation of a zonally uniform flow that affects the stationary wave response, as we are trying to isolate the acceleration due to stationary waves, without any background zonal flow. We will investigate the feedback of a zonal-mean flow on the stationary wave response in Section \ref{sec:jet-feedback-on-accn}. The $y$ direction is represented with a combination of sine and cosine functions (the ``Sin/Cos'' basis in \citet{burns2020dedalus}), which allows for even or odd parity to be imposed on the variables in this direction. The $u$, $w$, $\Phi$, and $T$ fields are required to be symmetric about the equator, and the $v$ field is required to be antisymmetric. The $z$ direction is represented with a Chebyshev basis, where we impose the boundary conditions $w=0$ at the top and bottom of the model.

This system of equations is similar to the primitive equations solved in a GCM; however, the solution is conceptually closer to the results of shallow-water models such as those in \citet{matsuno1966quasi}, \citet{showman2011superrotation}, and to the results of the three-dimensional stationary wave model of \citet{tsai2014three}. This system is on a beta-plane rather than a sphere, has strict parity conditions on all of its variables about the equator, is restricted to non-zero zonal wavenumbers, and is forced by a pre-defined stationary sinusoidal function. It therefore recovers only the three-dimensional stationary waves due to this forcing, rather than producing a full planetary circulation as a GCM would. The solution should be considered as ``pseudo-equilibrium'', as it satisfies the constrained set of equations we are solving, but in reality would produce a zonal-mean zonal flow which would modify the solution.

\subsection{Stationary Wave Structure}\label{sec:stationary-wave-structure}

\begin{figure*}
\centering 
\subfloat[With a linear Rayleigh drag]{%
  \includegraphics[width=\columnwidth]{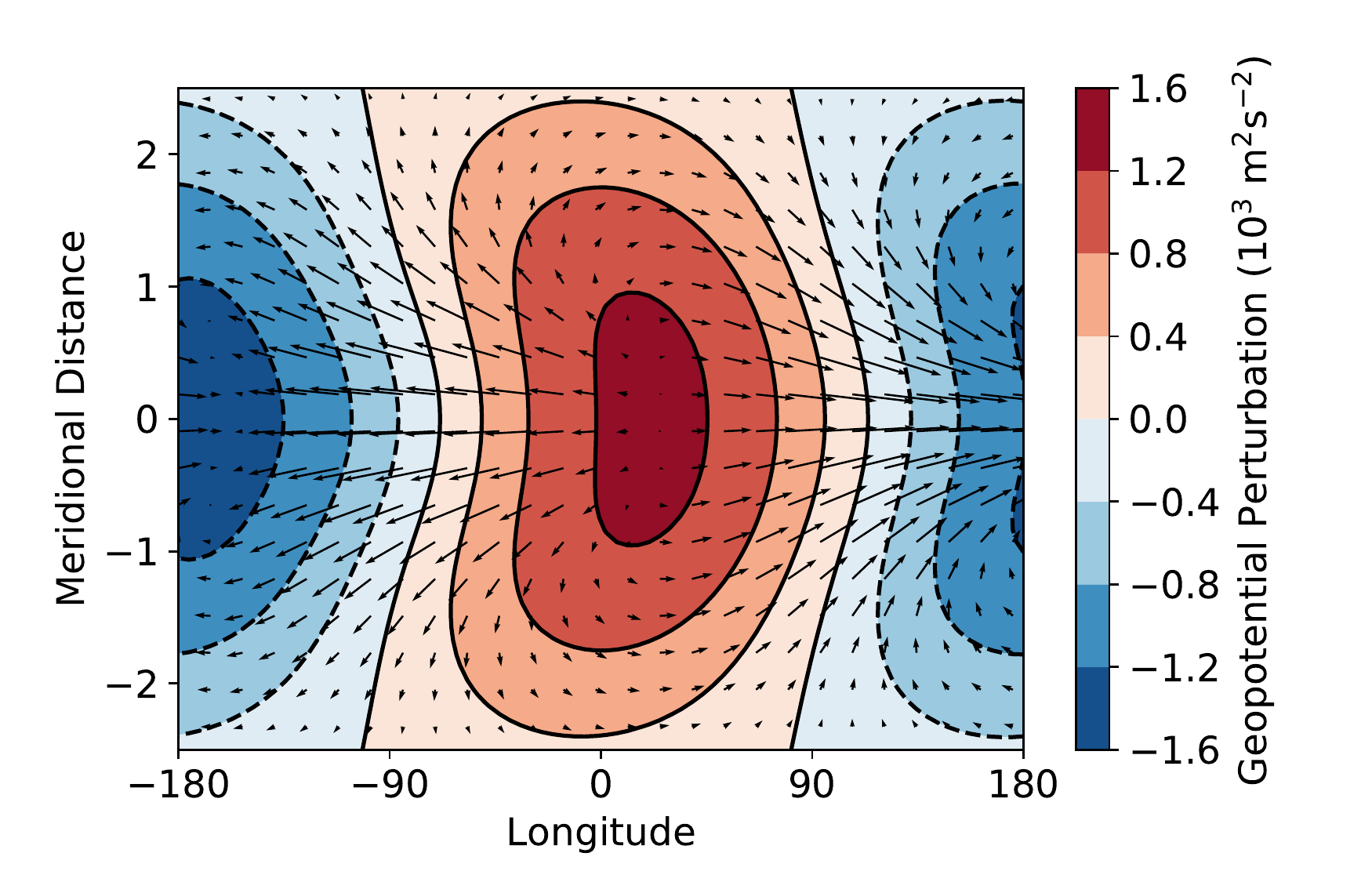}%
  \label{fig:T-map-quivers-lin}%
}\qquad
\subfloat[With no linear Rayleigh drag]{%
  \includegraphics[width=\columnwidth]{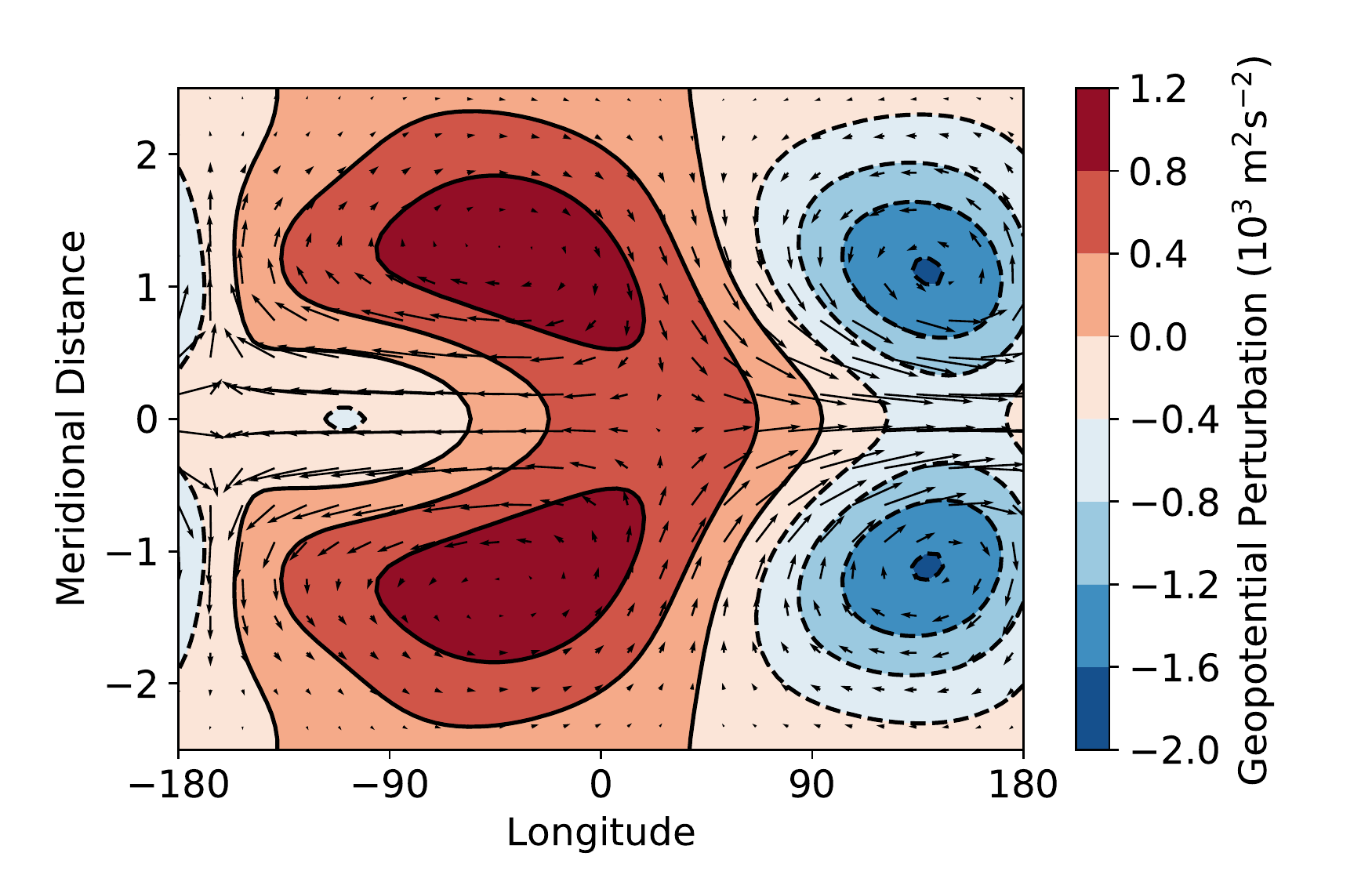}%
  \label{fig:T-map-quivers-nonlin}%
}
\caption{The geopotential and velocity fields at the $p=\SI{0.39}{\bar}$ ($z \approx \SI{8630}{\metre}$) level for the stationary solutions to Equation \ref{eqn:dedalus-equations}. This level corresponds to the peak of the ``Stationary Horizontal'' acceleration in Figure \ref{fig:accn-profiles-p}. The solution in the first plot has a linear drag $\alpha_{dyn}u$ for a drag timescale $\tau_{dyn} = 1/\alpha_{dyn} = 1 \si{\day}$, and the second plot has no linear drag. Away from the equator, where the momentum balance is governed by the Coriolis force, the two solutions are similar. Near the equator, the solutions are governed by linear drag and nonlinear balance respectively. This distinction is vital to accurately calculating the equatorial velocities and acceleration.}\label{fig:T-map-quivers-dedalus}
\end{figure*}

Previous studies have found the stationary wave response and resulting zonal acceleration in systems with a linear drag applied to the velocity fields \citep{tsai2014three}. \citet{showman2011superrotation} found that a linear shallow-water system with no linear drag produced no zonal acceleration at the equator, as it could not produce the required Kelvin wave response there without drag. In this study, we calculate the solution with no linear drag in order to match our GCM simulations, and to derive an acceleration and equatorial jet speed that does not depend on an arbitrarily chosen linear drag timescale. We will show that our non-linear model can produce a zonal acceleration at the equator without linear dynamical drag. \citet{showman2011superrotation} showed that this was possible in two-dimensional nonlinear shallow-water simulations, but did not examine the resulting stationary wave structure.

Figure \ref{fig:T-map-quivers-dedalus} shows the temperature and velocity fields of the equilibrated solutions to Equation \ref{eqn:dedalus-equations} in the Dedalus software package, with and without a linear drag $\alpha_{dyn}$. Both of the solutions have a planetary radius \SI{6e6}{\metre}, acceleration due to gravity \SI{10}{\metre\per\second\squared}, day (and year) length \SI{10}{\day}, instellation \SI{e3}{\watt\per\metre\squared}, specific heat capacity \SI{e3}{\joule\per\kilogram\per\kelvin}, molar mass \SI{28}{\gram\per\mol}, and surface pressure \SI{e5}{\pascal}. The square of the Brunt–Väisälä frequency is set to a constant value $N_{*}^{2} = \SI{5e-4}{\per\second\squared}$. The magnitude of the velocity field in the solution is sensitive to this parameter, but it is difficult to estimate accurately for a atmosphere in general (unless the atmosphere is isothermal, which is not the case here). We suggest that this value is appropriate for these ``Earth-like'' atmospheres, as it gives velocities that approximately match the GCM simulations. It is important to note that the overall magnitude of the acceleration fluxes is sensitive to this parameter, but if the actual mechanism is not.

\begin{figure*}
\centering 
\subfloat[With linear Rayleigh drag]{%
  \includegraphics[width=\columnwidth]{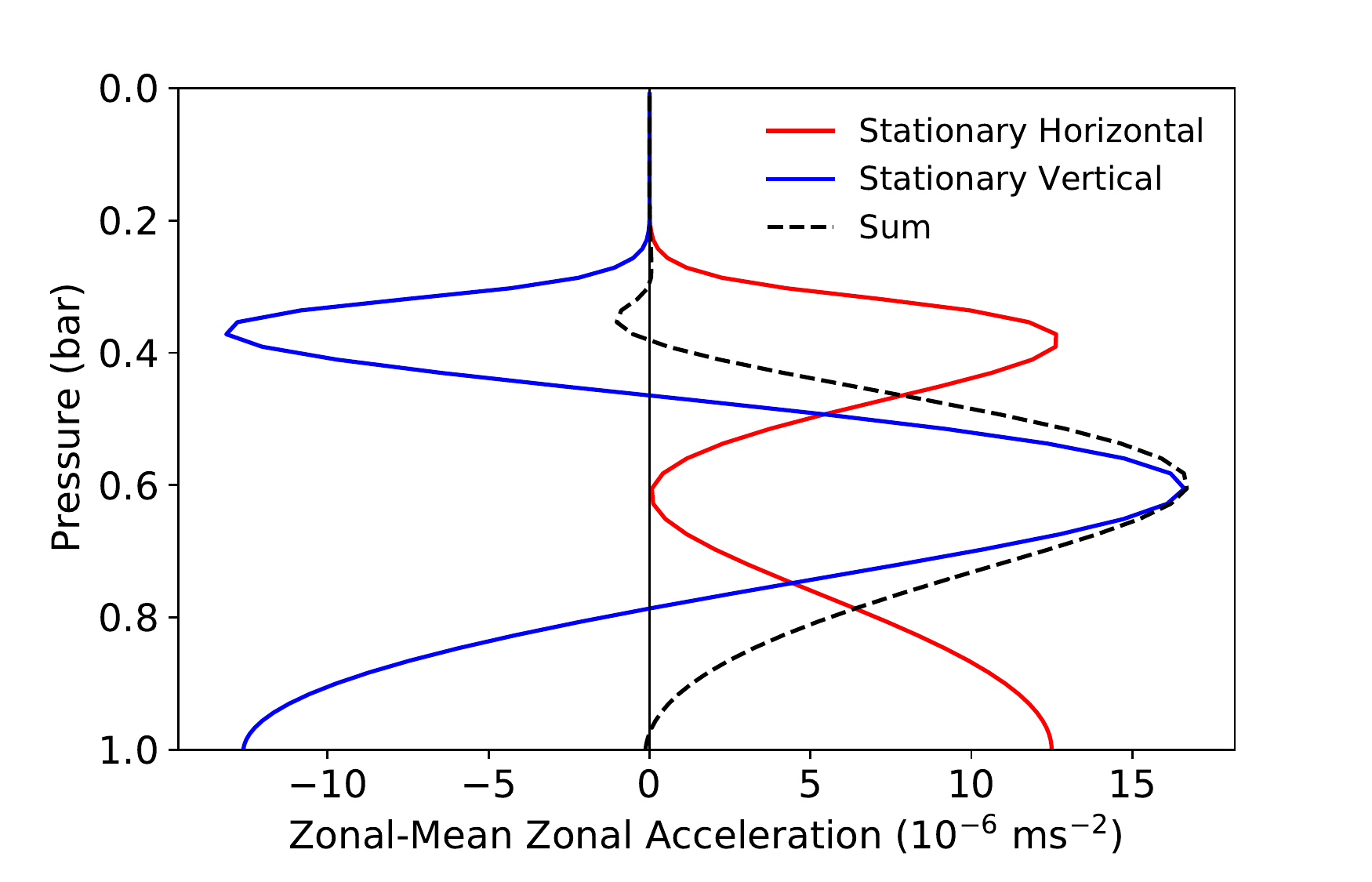}%
  \label{fig:accn-profiles-p-lin}%
}\qquad
\subfloat[Without linear Rayleigh drag]{%
  \includegraphics[width=\columnwidth]{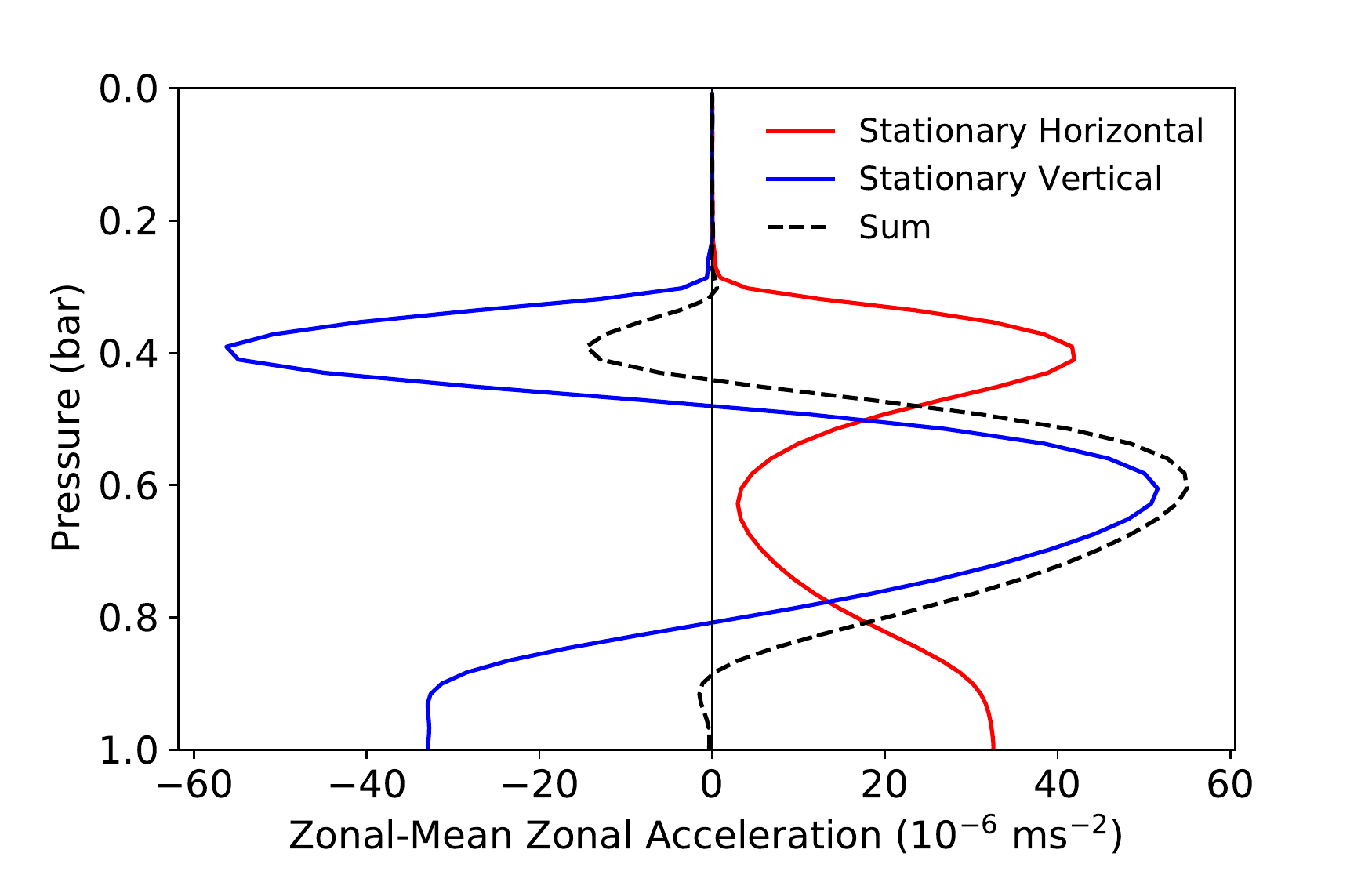}%
  \label{fig:accn-profiles-p-nonlin}%
}
\caption{The vertical profiles of the ``stationary'' acceleration terms in Equation \ref{eqn:zonal-mean-acceleration-reduced} for the solutions in Figure \ref{fig:T-map-quivers-dedalus}, which are the basis of the mechanism in Section \ref{sec:jet-formation-mechanism}. The qualitative forms of the profiles are the same, but their magnitudes are different as the zonal momentum equation is governed by different balances in each case. The local peaks of the acceleration profiles at about \SI{0.4}{\bar} correspond to the acceleration in two-dimensional shallow-water models such as in \citet{showman2011superrotation}. The peak of the ``Stationary Vertical'' term at about \SI{0.6}{\bar} is the main contribution to the unbalanced initial acceleration, which produces the equatorial jet.}\label{fig:accn-profiles-p}
\end{figure*}

These parameters correspond approximately to the GCM test in Section \ref{sec:gcm-simulation-results} with instellation \SI{e3}{\watt\per\metre\squared}. Figure \ref{fig:T-map-quivers-lin} shows the ``linear'' solution with a strong linear dynamical damping with a timescale of \SI{1}{\day}. While gaseous planets such as hot Jupiters may have be affected by magnetic drag that can be represented by this term \citep{komacek2016daynighti}, there is no reason to expect that the equatorial jet speed of terrestrial planets at the temperatures we consider will be affected be a uniform Rayleigh drag (apart from the drag near their surface, which will not affect the momentum budget at the level of the jet). We include the case with linear drag for comparison with the two-dimensional shallow-water solutions of \citet{matsuno1966quasi} and \citet{showman2011superrotation}, as well as the three-dimensional stationary wave calculations of \citet{tsai2014three}. The geopotential and velocity fields in Figure \ref{fig:T-map-quivers-lin} are similar to those in \citet{matsuno1966quasi} and \citet{showman2011superrotation}, with a stronger response on the equator than the solution in \citet{matsuno1966quasi} due to the higher radiative and dynamical damping rates.

Figure \ref{fig:T-map-quivers-nonlin} shows the ``non-linear'' solution with the same parameters as the solution in Figure \ref{fig:T-map-quivers-lin}, but with no dynamical damping. This system is governed by the same balances in the momentum and thermodynamic equations in Equation \ref{eqn:dedalus-equations} as the GCM simulations later. \citet{komacek2016daynighti} discusses this non-linear (or ``advective'') momentum balance in detail, and compares it to other momentum balances via the Coriolis term and linear damping. The `linear'' and ``non-linear'' solutions are similar far away from the equator, where the momentum balance is governed by the Coriolis term $f \times \boldsymbol{u}$. 

In the linear case, the geopotential gradient $\nabla_{z} \Phi$ is balanced by the linear drag $\alpha_{dyn}\boldsymbol{u}$. This leads to a linear scaling between the velocity and the geopotential perturbation (and, via the hydrostatic balance equation and the thermodynamic equation, between the velocity and the forcing):

\begin{equation}
    u \sim \Delta \Phi \sim \Delta T \sim Q.
\end{equation}

This gives the linear relation between the velocity perturbations and forcing in \citet{matsuno1966quasi} and \citet{showman2011superrotation}, which means that the zonal acceleration caused by these perturbations scales quadratically with the forcing. In the ``non-linear'' case this Rayleigh drag term is not present and the balance must be different. We discuss this non-linear balance in more detail in Section \ref{sec:predicting-jet-speed}, and show how it leads to a much weaker dependence of the velocity perturbations and zonal acceleration on the magnitude of the stellar forcing. This approximation will be invalid when the WTG approximation does not apply globally on sufficiently rapidly rotating planets, or for sufficiently strong forcing that leads to non-negligible non-linear terms in the thermodynamic equation in Equation \ref{eqn:dedalus-equations} \citep{pierrehumbert2018review}.

An important difference between the ``linear'' and ``non-linear'' cases is that in the linear case the magnitude of the velocity field is entirely dependent on the arbitrarily chosen $\alpha_{dyn}$ damping parameter. This means that for a given forcing $Q$ (and constant planetary parameters), the magnitude of the equatorial acceleration and the resulting jet speed will be entirely determined by the choice of the dynamical damping rate $\alpha_{dyn}$. Conversely, in the ``non-linear'' case, the magnitude of the equatorial velocities is determined only by the planetary parameters. The non-linear term plays the role of the linear dynamical damping, balancing the gradient of the geopotential $\Phi$. The non-linear solution therefore has more predictive power than the linear solution. The linear case is still useful for emulating the behaviour of the non-linear case, and providing analytically soluble systems, as the damping parameter can represent the non-linear term if an appropriate value is chosen (for example, if $\alpha_{dyn} u \sim u \partial_{x}u$).

Both solutions are sensitive to the Brunt–Väisälä frequency $N_{*}$. If the frequency is large enough that the $w$ term is the dominant balance in the thermodynamic equation in Equation \ref{eqn:dedalus-equations} (which is the case in the solutions plotted in this study), the qualitative form of the solutions should not depend on the exact magnitude of  $N_{*}$. The absolute magnitude of the vertical velocity perturbation (and by extension, the horizontal velocity perturbations) will depend on its magnitude. If the Brunt–Väisälä frequency is small enough that the $w$ term is not the main term balancing the forcing in the thermodynamic equation (i.e. the WTG approximation does not apply), the stationary wave response to forcing may be very different. The Brunt–Väisälä frequency depends on the temperature structure of the atmosphere; it is simple to calculate for an isothermal atmosphere but we found that assuming an isothermal atmosphere did not produce stationary wave solutions that matched the GCM simulations, where the convective atmosphere (on the day-side) is much closer to neutral stability. The Brunt–Väisälä frequency is zero for a neutrally stable atmosphere, and it is difficult to estimate an accurate value to represent the entire atmosphere of a tidally locked planet. 

We chose a constant value of $N_{*}^{2} = \SI{5e-4}{\per\second\squared}$ to give solutions that approximately matched the magnitude of the stationary waves in the GCM, and to be consistent with the value used to represent the tropics of the Earth by \citet{wu2000vertical}. The absolute value of the jet speed we predict is only as accurate as the value chosen for this frequency, although the qualitative mechanism is the same whatever value is chosen. An improved representation of this frequency could be a way to improve the accuracy of this model, highlighting the importance of understanding the generation of static stability in the atmospheres of tidally locked planets.

\begin{figure*}
\centering 
\subfloat[Evolution of the zonal-mean zonal velocity]{%
  \includegraphics[width=\columnwidth]{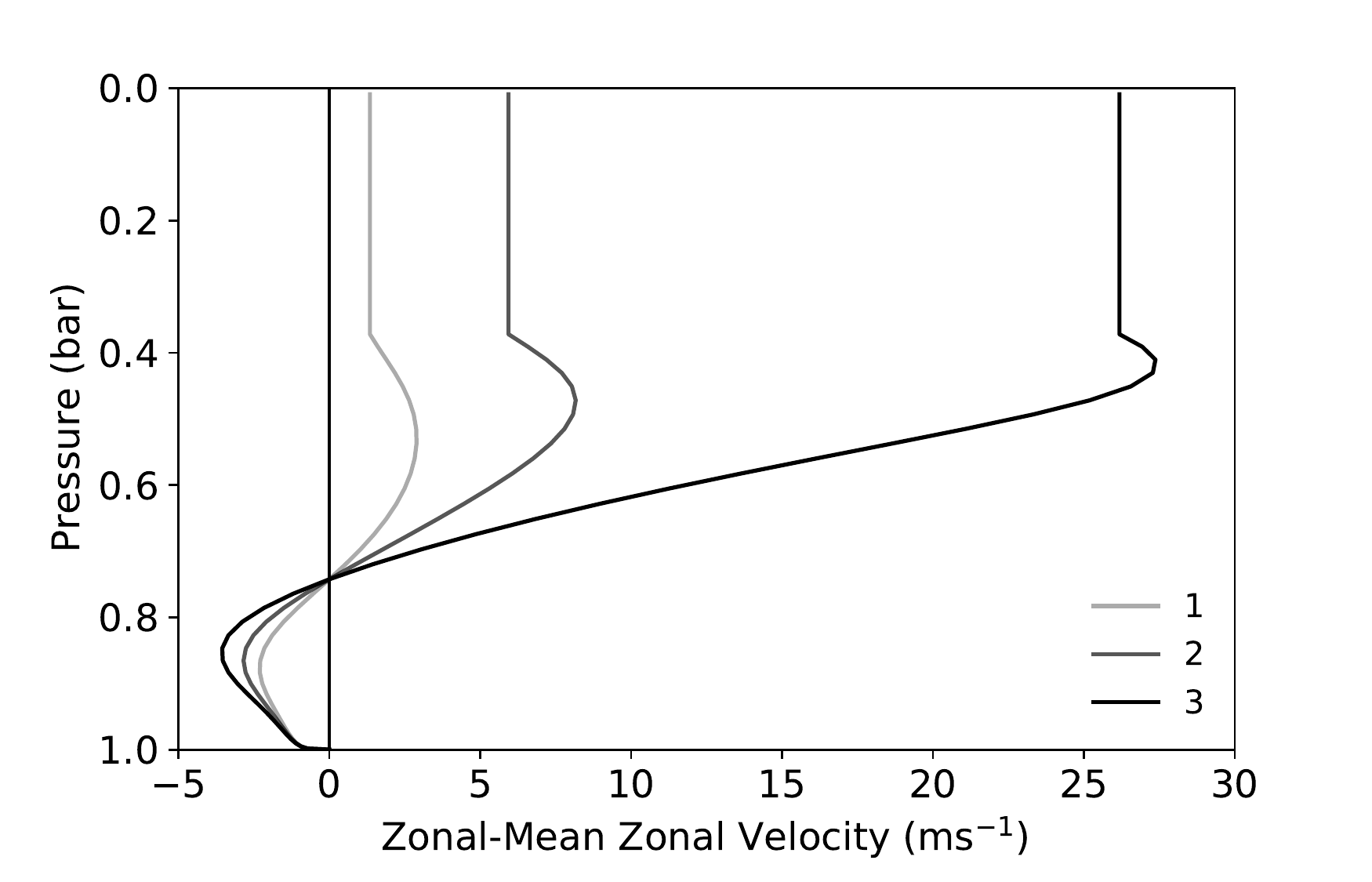}%
  \label{fig:accn_schematic_ubar}%
}\qquad
\subfloat[Evolution of the zonal momentum budget]{%
  \includegraphics[width=\columnwidth]{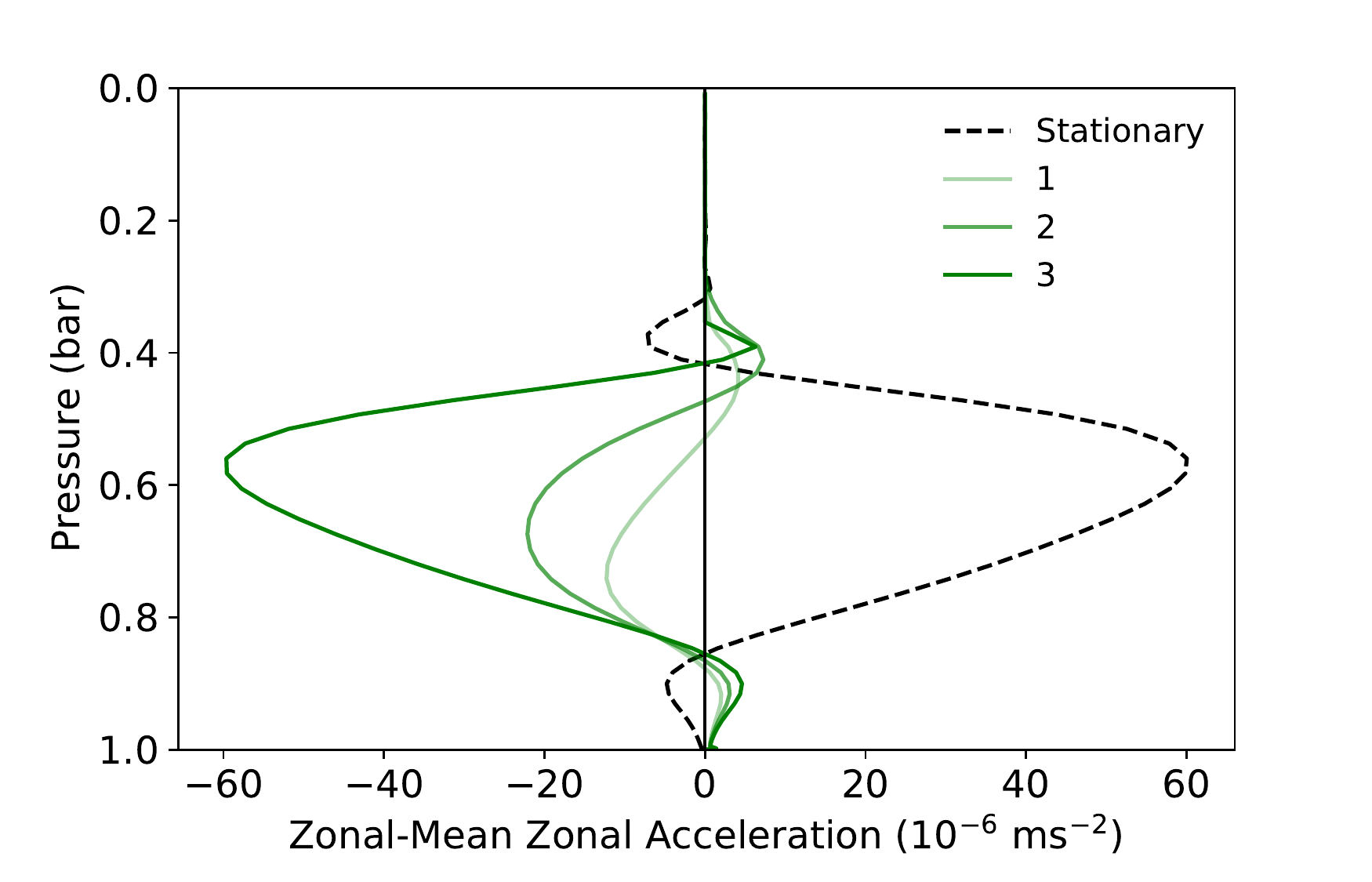}%
  \label{fig:accn_schematic}%
}
\caption{A schematic of the formation of an equatorial jet by the mechanism proposed in this study. The first panel represents the development of the jet; line 3 is the equilibrium jet predicted by Equation \ref{eqn:jet-speed-estimate} to balance the ``Stationary'' acceleration terms, which are shown by the dashed line in the second panel. It produces line 3 in the second panel, which balances the ``Stationary'' terms. Lines 1 and 2 in the first panel are velocity profiles chosen to represent the jet in its spin-up; as time progresses, the jet speed increases and the ``Mean Vertical'' acceleration in the second panel increases as well, until it reaches line 3 and equilibrium is achieved.}\label{fig:accn-schematic}
\end{figure*}

\section{Jet Formation Mechanism}\label{sec:jet-formation-mechanism}

The three-dimensional stationary waves induced by the day-night instellation gradient produce a zonal acceleration at the equator (which would accelerate an equatorial jet, if we had not suppressed this in our idealised calculations). In this section, we calculate the vertical profiles of the different terms contributing to this acceleration. We will then propose a mechanism in which the jet forms due to this acceleration, then interacts with the vertical velocity to produce a deceleration that balances this acceleration and produces equilibrium. This mechanism is similar to that used to describe the formation of superrotating flows on Venus and Titan by \citet{leovy1987zonal} and \citet{zhu2006maintenance}.

\subsection{Acceleration Profile Structure}

The velocity fields in the solution to Equation \ref{eqn:dedalus-equations} produce a zonal-mean zonal acceleration in spherical pressure coordinates $(\theta,\phi,p)$ \citep{lutsko2018response}:

\begin{equation}\label{eqn:zonal-mean-acceleration-unsimplified}
  \begin{aligned} \frac{\partial \overline{u}}{\partial t}=& 
  \underbrace{f\overline{v} - \frac{\overline{v}}{a \cos ^{2} \phi} \frac{\partial}{\partial \phi}\left(\overline{u}  \cos ^{2} \phi\right)}_{\mathrm{Term\ MH}} \underbrace{-\overline{\omega}\frac{\partial \overline{u}}{\partial p}}_{\mathrm{Term\ MV}}\\
    &\underbrace{- \frac{1}{a \cos ^{2} \phi} \frac{\partial}{\partial \phi}\left(\overline{u^{*} v^{*}} \cos ^{2} \phi\right)}_{\mathrm{Term\ SH}} \underbrace{-\frac{\partial}{\partial p}\left(\overline{u^{*} \omega^{*}}\right)}_{\mathrm{Term\ SV}} \\
    &\underbrace{-\frac{1}{a \cos ^{2} \phi} \frac{\partial}{\partial \phi}\left([\overline{u^{\prime} v^{\prime}}] \cos ^{2} \phi\right)}_{\mathrm{Term\ TH}} \underbrace{-\frac{\partial}{\partial p}[\overline{u^{\prime} \omega^{\prime}}]}_{\mathrm{Term\ TV}}, \end{aligned}
\end{equation}

for the acceleration terms ``Mean Horizontal'' (MH), ``Mean Vertical'' (MV), ``Stationary Horizontal'' (SH), ``Stationary Vertical'' (SV), ``Transient Horizontal'' (TH), and ``Transient Vertical'' (TV). On the beta-plane $(x,y,p)$ this is:

\begin{equation}\label{eqn:zonal-mean-acceleration-y}
  \begin{aligned} \frac{\partial \overline{u}}{\partial t}=& 
  \underbrace{f\overline{v} - \overline{v}\frac{\partial \overline{u}}{\partial y}}_{\mathrm{Term\ MH}} \underbrace{-\overline{\omega}\frac{\partial \overline{u}}{\partial p}}_{\mathrm{Term\ MV}}\\
    &\underbrace{ \frac{\partial}{\partial y}\left(\overline{u}^{*} \overline{v}^{*}\right) }_{\mathrm{Term\ SH}} \underbrace{-\frac{\partial}{\partial p}\left(\overline{u}^{*} \overline{\omega}^{*}\right)}_{\mathrm{Term\ SV}} \\
    &\underbrace{ \frac{\partial}{\partial y}[\overline{u^{\prime} v^{\prime}}]}_{\mathrm{Term\ TH}} \underbrace{-\frac{\partial}{\partial p}[\overline{u^{\prime} \omega^{\prime}}]}_{\mathrm{Term\ TV}}, \end{aligned}
\end{equation}

where overbars denote zonal-mean values, and asterisks denote deviations from the zonal-mean value (``stationary'' terms). All quantities such as $u$ are time-means, apart from the ``Transient'' terms where the primes denote deviations from the time-mean (``eddy'' terms), and the square brackets denote a time-mean taken after the two eddy terms are multiplied together. Note that these terms are simplified compared to the equivalent terms in other studies such as \citet{mayne2017hotjupiter}; the ``Mean'' terms have been expanded and partially cancelled by combination with the zonally-averaged continuity equation.

For forcing that is symmetric about the equator, $\overline{v}$ will be zero on the equator. We assume that the stationary forcing will produce stationary waves that are much larger than the transient waves, so the acceleration terms associated with the transient waves can be neglected (although we will still include these terms when diagnosing the momentum budget in the GCM simulations). These assumptions lead to the following simplified expression for the zonal-mean zonal acceleration on the equator:

\begin{equation}\label{eqn:zonal-mean-acceleration-reduced}
  \begin{aligned} \frac{\partial \overline{u}}{\partial t}=& \underbrace{- \overline{\omega} \frac{\partial\overline{u}}{\partial p}}_{\mathrm{Term\ MV}} \underbrace{-\frac{\partial}{\partial p}\left(\overline{u^{*} \omega^{*}}\right)}_{\mathrm{Term\ SV}} \\
    &\underbrace{- \frac{1}{a \cos ^{2} \phi} \frac{\partial}{\partial \phi}\left(\overline{u^{*} v^{*}} \cos ^{2} \phi\right)}_{\mathrm{Term\ SH}}.  \\
 \end{aligned}
\end{equation}

Figure \ref{fig:accn-profiles-p} shows the terms in Equation \ref{eqn:zonal-mean-acceleration-reduced}, for the solution to Equation \ref{eqn:dedalus-equations} in the linear limit and the nonlinear limit (which are the solutions shown in Figure \ref{fig:T-map-quivers-dedalus}). The ``Mean Vertical'' term is zero in both cases as the zonal-mean zonal velocity is zero. The linear and non-linear cases have qualitatively similar vertical acceleration profiles, although the non-linear case has a larger acceleration, as the linear case must have smaller velocity perturbations from its additional dynamical damping. The ``Stationary Horizontal'' term corresponds to a transport of eastward momentum towards the equator at about \SI{0.4}{\bar}, as shown by \citet{showman2011superrotation} in a two-dimensional model corresponding to this level. The ``Stationary Vertical'' term corresponds  to a transport of eastward momentum down from this level to about \SI{0.6}{\bar}, where it will accelerate the initial jet at the initialisation of the atmosphere from rest in the GCM. The structure of this acceleration is similar to that shown by \citet{debras2020acceleration} for hot Jupiters, who identified the role of the ``Stationary Vertical'' term in accelerating the jet, rather than only decelerating it as in \citet{showman2011superrotation}. In the next section, we will describe a mechanism where this initial jet produces a new ``Mean Vertical'' acceleration term, which moves it up back towards the \SI{0.4}{\bar} level, and eventually balances the ``Stationary'' terms in equilibrium.

\subsection{Predicted Jet Speed}

\begin{figure*}
\centering 
\subfloat[Equilibrium jet profiles for different instellations]{%
  \includegraphics[width=\columnwidth]{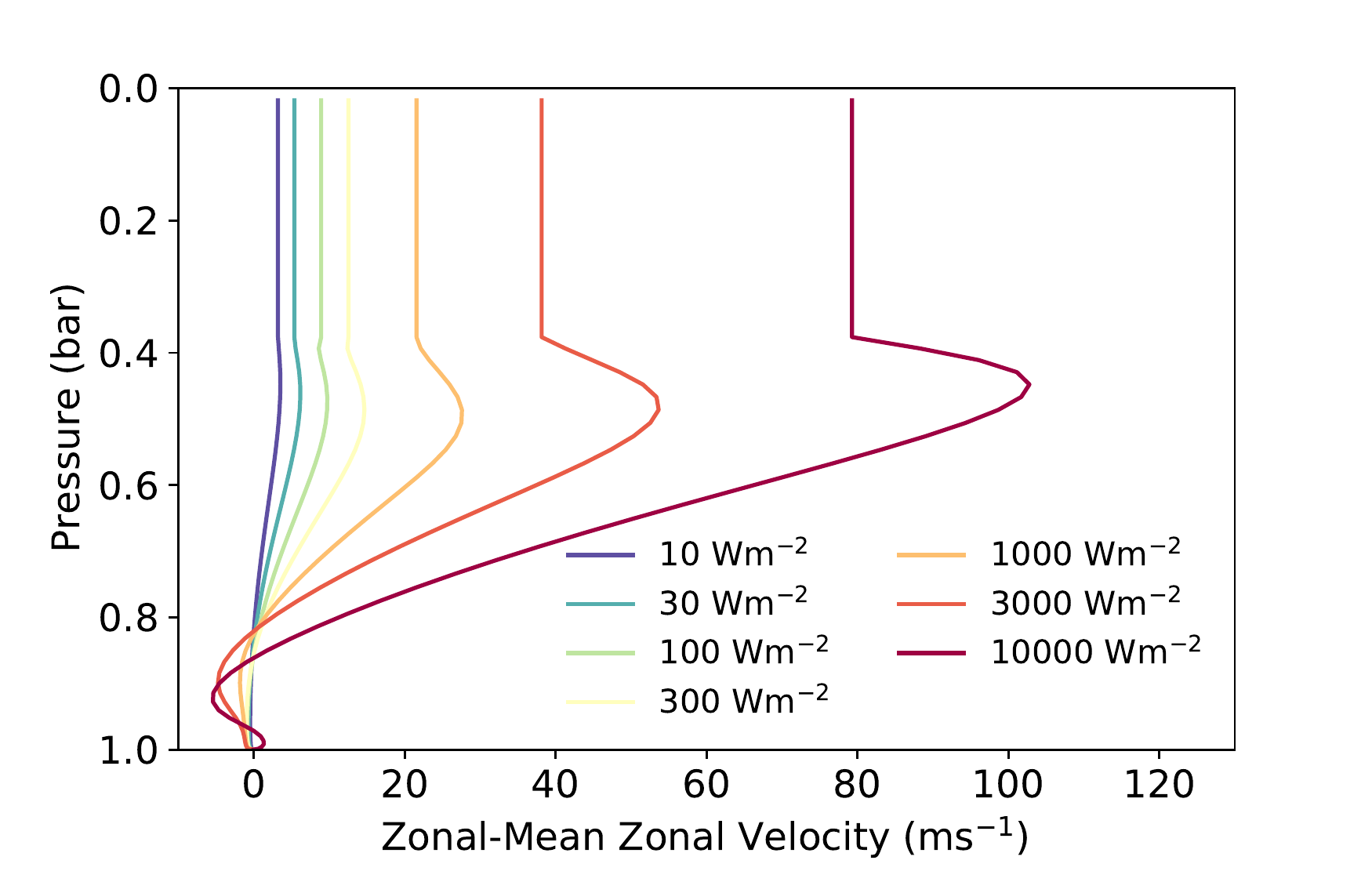}%
  \label{fig:Ubar-scaling-dedalus-profiles}%
}\qquad
\subfloat[Maximum jet speed versus instellation]{%
  \includegraphics[width=\columnwidth]{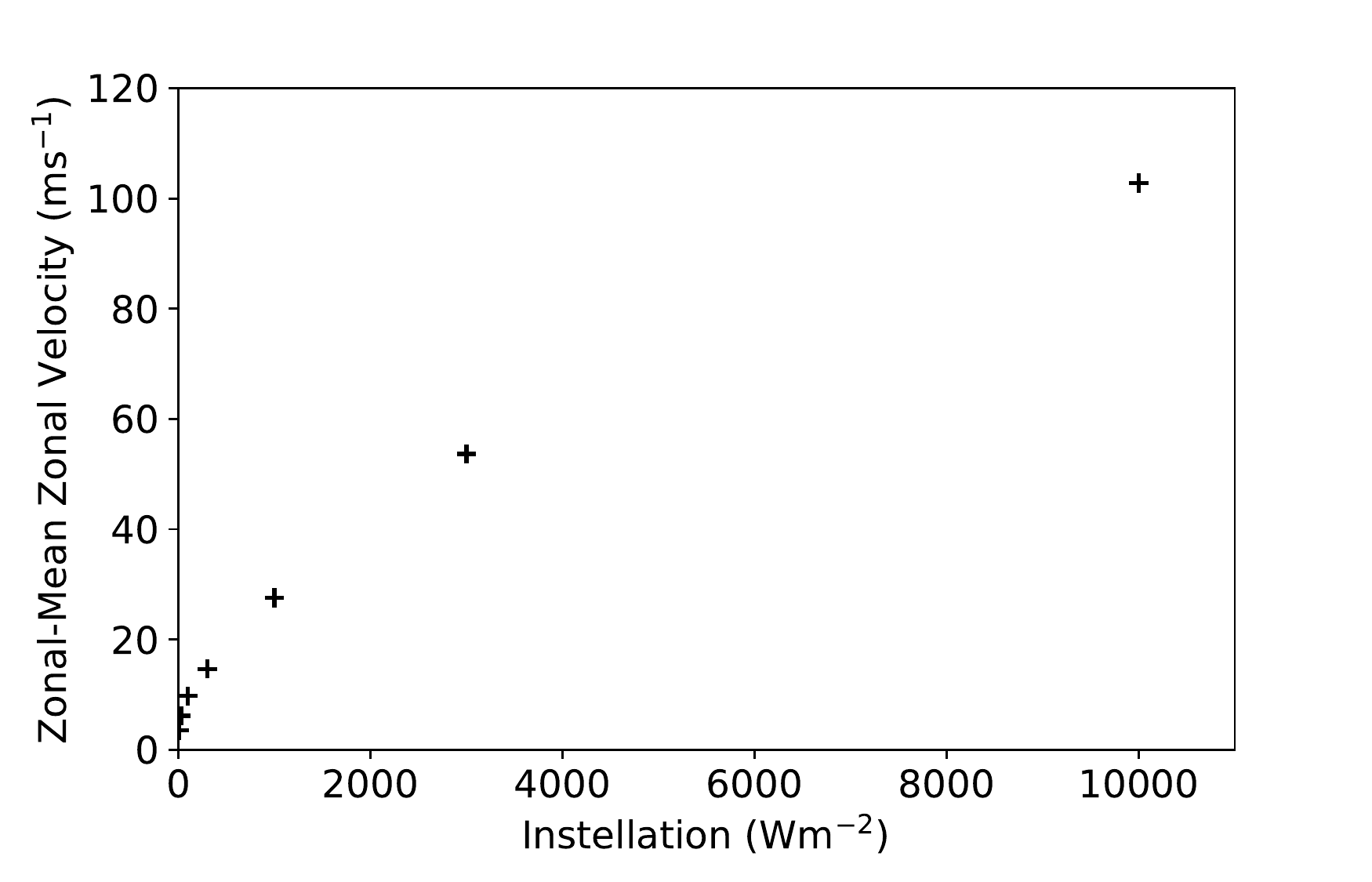}%
  \label{fig:Ubar-scaling-dedalus-nonlin}%
}
\caption{Equilibrium zonal-mean zonal velocity profiles predicted by the ``non-linear'' (zero Rayleigh drag) calculation using the Dedalus software described in Section \ref{sec:3d-stationary-wave-response}, according to the mechanism in Section \ref{sec:jet-formation-mechanism}. The first panel shows the profiles for different values of instellation, and the second panel shows the maximum value of each profile versus the instellation. The jet speed only depends weakly on instellation, due to the non-linear balance governing the magnitude of the velocity perturbations at the equator, its inverse dependence on the zonal-mean vertical velocity.}\label{fig:ubar-profiles-dedalus}
\end{figure*}

The acceleration profiles caused by momentum transport from stationary waves in Figure \ref{fig:accn-profiles-p} correspond to the zonal acceleration when there is no zonal-mean zonal flow. We suppose in this section that the jet does not strongly feed back on these stationary waves and affect the resulting acceleration, an assumption that we test in Section \ref{sec:jet-feedback-on-accn}. Instead, the primary effect of the zonal-mean jet $\overline{u}$ is to increase the ``Mean Vertical'' term in Equation \ref{eqn:zonal-mean-acceleration-reduced}, which is zero when the atmosphere is at rest. If this is the only feedback from the jet on the zonal momentum budget, the jet will accelerate until the ``Mean Vertical'' term exactly balances the sum of the ``Stationary'' terms, shown as the dashed black line in Figure \ref{fig:accn-profiles-p}.

Setting $\frac{\partial \overline{u}}{\partial t} = 0$ in Equation \ref{eqn:zonal-mean-acceleration-reduced} shows that the zonal-mean jet $\overline{u}$ required for the ``Mean Vertical'' term to balance the ``Stationary'' terms is:

\begin{equation}\label{eqn:predicted-jet-speed}
  \begin{aligned} \overline{u} =& \int - \frac{1}{\overline{\omega}} \left( \frac{1}{a \cos ^{2} \phi} \frac{\partial}{\partial \phi}\left(\left(\overline{u^{*} v^{*}}\right) \cos ^{2} \phi\right) -\frac{\partial}{\partial p}\left(\overline{u^{*} \omega^{*}}\right)\right) dp .
 \end{aligned}
\end{equation}

Figure \ref{fig:accn-schematic} is a schematic of how this mechanism works in practice. The first panel shows the evolution of the zonal-mean zonal flow to equilibrium, where line 3 is the equilibrium jet predicted by Equation \ref{eqn:predicted-jet-speed}, and lines 1 and 2 are example jet profiles chosen to represent the spin-up of the jet. The second panel shows the resulting ``Mean Vertical'' acceleration from each of these velocity profiles. As the jet speed increases over time, this term increases until it balances the sum of the ``Stationary Horizontal'' and ``Stationary Vertical'' terms in the zonal-mean momentum equation and equilibrium is reached. Line 3 and the dashed line do not cancel exactly above $z=H$ due to the way we estimate $\overline{\omega}$, which we will discuss later.

The mechanism can also be understood qualitatively by considering the direction of zonal momentum transport. The stationary waves form in response to the day-side instellation and night-side cooling. These produce a ``Stationary Horizontal'' transport of eastward angular momentum towards the equator, with a peak around one scale height above the surface. This is opposed by the ``Stationary Vertical'' transport of this eastward momentum towards the surface. This accelerates a jet closer to the surface than the peak of the horizontal momentum transport. As this jet forms it interacts with the zonal-mean vertical velocity associated with a Hadley-like circulation (which is not zero, as there is a zonal-mean component to the instellation, unlike in the idealised sinusoidal forcing in \citet{showman2011superrotation}). This interaction produces a ``Mean Vertical'' accleration, as the jet is moved upwards by the zonal-mean vertical velocity. As the jet increases, this acceleration term balances the ``Stationary'' terms, until they are exactly balanced in equilibrium. If the ``Stationary'' terms exactly cancel at a certain pressure level, the peak of the equilibrium jet will form there, as the ``Mean Vertical'' term will be zero there as the jet has no vertical shear at that point.

The terms denoted by asterisks in Equation \ref{eqn:predicted-jet-speed} are determined entirely by the calculation of the stationary wave response in Section \ref{sec:3d-stationary-wave-response}. This calculation specifically excludes the zonal-mean quantities with zonal wavenumber 0, so does not determine the zonal-mean vertical velocity $\overline{w}$. We instead estimate $\overline{w}$ to be the zonal mean of the day-side vertical velocity, as we expect there to be a Hadley-like overturning on the day-side due to instellation there. The sinusoidal variation of forcing with longitude in Section \ref{sec:3d-stationary-wave-response} implies an equal and opposite overturning on the night-side, giving no zonal-mean vertical velocity. A more realistic forcing field like that in \citet{perez2013atmospheric} would be uniform on the night-side, and would only support overturning on the day-side, giving a non-zero zonal-mean vertical velocity. Our approximation of $\overline{w}$ in this stationary wave calculation is therefore:

\begin{equation}
    \overline{\omega}(y,p) = \frac{1}{\pi} \int_{-\pi/2}^{\pi/2} \omega(x,y,p) dx.
\end{equation}

We only calculate $\overline{u}$ up to $z=H$ using this approximation, as above this level  $\overline{\omega}$ can cross zero, giving an undefined solution to Equation \ref{eqn:predicted-jet-speed}. Figure \ref{fig:accn_schematic} shows that neglecting $\overline{u}$ above $z=H$ is a reasonable approximation, as the majority of the ``Stationary'' acceleration is below this level, so the portion of $\overline{u}$ above $z=H$ is not important to the zonal-mean momentum budget.

Figure \ref{fig:ubar-profiles-dedalus} shows the zonal velocity profiles that are predicted by this mechanism using the stationary wave calculations in Dedalus, as discussed in Section \ref{sec:3d-stationary-wave-response}. We vary the instellation (the magnitude of the forcing) but keep all other parameters the same. It is notable that the predicted zonal-mean jet only depends weakly on instellation. This is due to the nonlinear dependence of the velocity perturbations on the instellation, as well as the inverse relation between the jet speed and the zonal-mean vertical velocity (which increases with increasing instellation) given by Equation \ref{eqn:predicted-jet-speed}. In reality, and in the GCM simulations, the evolution of the global circulation will be more complicated as the jet will affect the ``Stationary'' terms and be affected by the ``Transient'' terms, among other non-linear processes. In the next section, we investigate how strongly the jet affects the ``Stationary'' terms.

\subsection{Jet Feedback on Stationary Acceleration Terms}\label{sec:jet-feedback-on-accn}

\begin{figure*}
\centering 
\subfloat[With an imposed equatorial jet \SI{30}{\metre\per\second}]{%
  \includegraphics[width=\columnwidth]{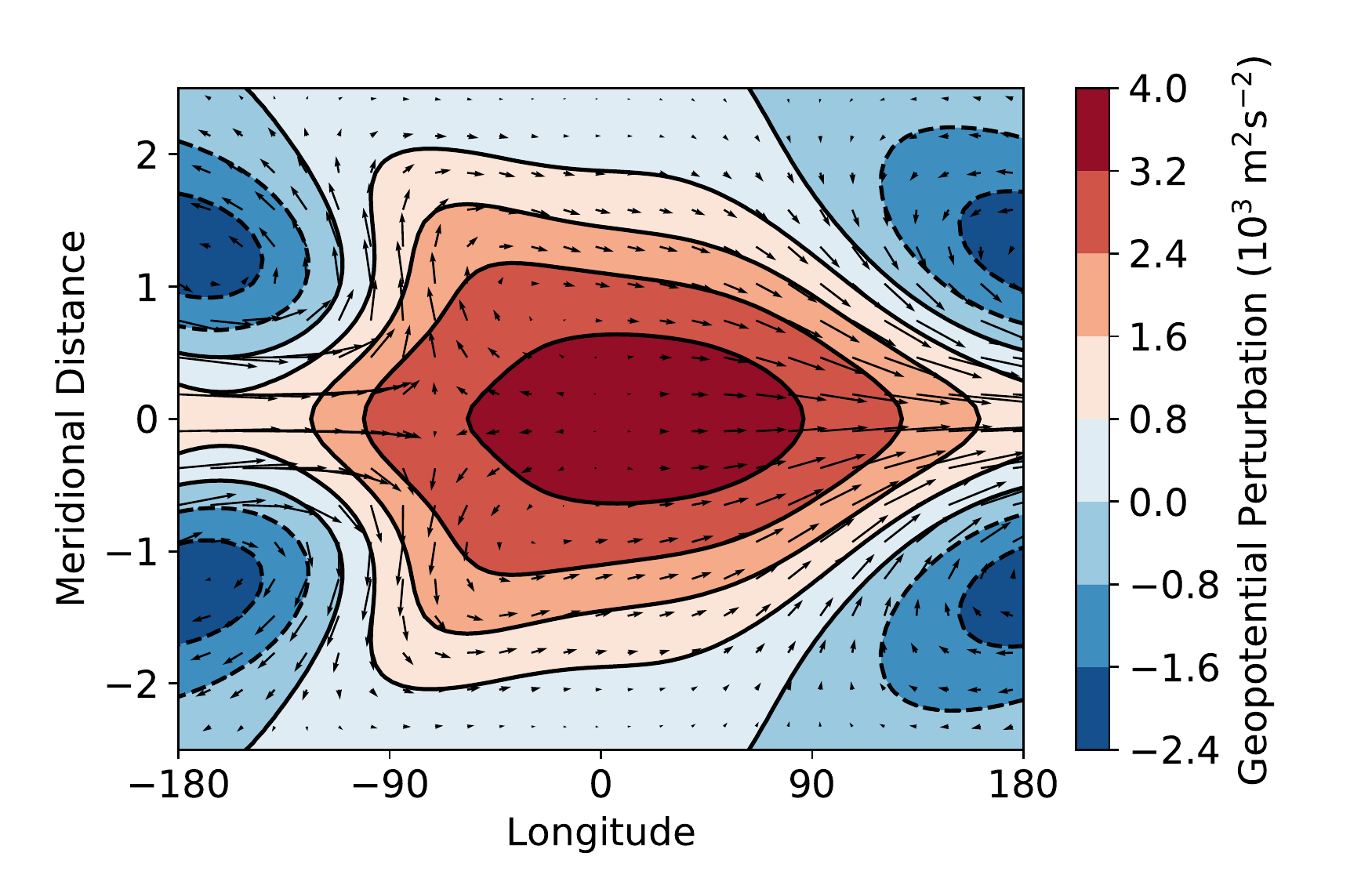}%
  \label{fig:phi-map-quivers-speed-30}%
}\qquad
\subfloat[Fractional ``Stationary Horizontal'' term versus imposed jet]{%
  \includegraphics[width=\columnwidth]{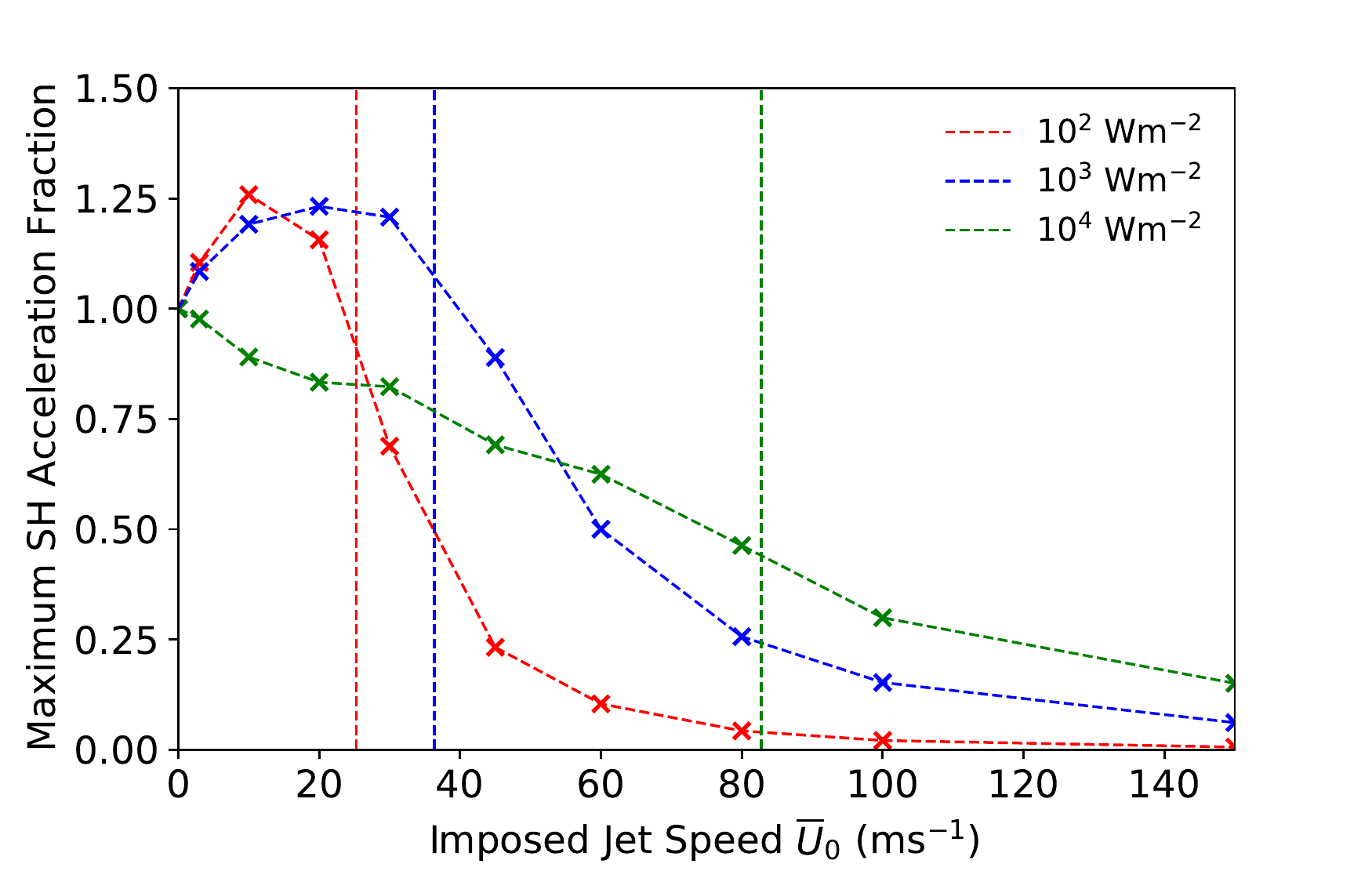}%
  \label{fig:jet-accn-feedback}%
}
\caption{The first panel shows the stationary wave response to forcing given an instellation \SI{e3}{\watt\per\metre\squared} and a background jet $\overline{u} = \overline{u}_{0} e^{-(y/a)^{2}}$, where $\overline{u}_{0} = \SI{30}{\metre\per\second}$ (similar to the equivalent test in the GCM). The geopotential and velocity fields appear very different to the solution in Figure \ref{fig:T-map-quivers-nonlin}, which is the same apart from the imposed background jet. This case with a background jet has almost exactly the same zonal acceleration at the equator, as shown by the second panel, which plots the fractional change in the maximum ``Stationary Horizontal'' acceleration relative to the acceleration with zero zonal-mean background flow. The vertical dashed lines show the maximum jet speed in the corresponding GCM simulation.}\label{fig:jet-feedback-plots}
\end{figure*}

This study proposes that the jet on terrestrial tidally locked planets reaches equilibrium when it becomes strong enough that the ``Mean Vertical'' acceleration term cancels the ``Stationary'' acceleration terms in Equation \ref{eqn:zonal-mean-acceleration-reduced}. \citet{tsai2014three} proposed a different mechanism for equilibrium in the atmospheres of hot Jupiters, where the jet Doppler-shifts the stationary waves produced by the instellation far enough eastward, that the ``Stationary'' acceleration terms decrease to zero. These acceleration terms require a phase difference between the on-equator stationary Kelvin waves and the off-equator stationary Rossby waves; if they are shifted towards the same position the acceleration will decrease \citep{tsai2014three,hammond2018wavemean}.

So far, we have neglected this effect and assumed that the jet does not affect the stationary waves and the resulting acceleration strongly enough to affect the zonal momentum budget. In this section, we find the effect of an imposed background flow on the non-linear solutions to Equation \ref{eqn:dedalus-equations}, to see if this feedback is an important effect. In Section \ref{sec:3d-stationary-wave-response} the zonal mean of all variables was set to zero, but we now impose a non-zero zonal-mean background flow with a Gaussian profile in the meridional direction, similar to the background flow in \citet{hammond2018wavemean}:

\begin{equation}
    \overline{u} = \overline{u}_{0} e^{-(\frac{y}{a})^{2}},
\end{equation}

where $a$ is the planetary radius. The exact meridional profile of this flow is not critical, as we focus on its effect at the equator only. The background flow must satisfy Equation \ref{eqn:dedalus-equations}, so we also impose a geopotential perturbation:

\begin{equation}
    \overline{\Phi} = a^{2} \beta  \overline{u}_{0} e^{-(\frac{y}{a})^{2}}.
\end{equation}

We impose the same zonal flow at all vertical levels; in reality the jet would vary in the vertical direction.

Figure \ref{fig:phi-map-quivers-speed-30} shows the geopotential and velocity fields for the non-linear solution with this background jet now imposed, with otherwise the same parameters as the solution with instellation \SI{1000}{\watt\per\metre\squared} in Figure \ref{fig:T-map-quivers-nonlin}. The imposition of the jet gives a visually different solution (with a structure explained in \citet{hammond2018wavemean}, but the magnitude of the zonal acceleration is almost the same in both cases. This shows that it can be difficult to estimate the zonal acceleration from the visual appearance of the stationary wave field.

Figure \ref{fig:jet-accn-feedback} shows the effect of the zonal-mean zonal velocity on the maximum value of the ``Stationary Horizontal'' acceleration term, for a range of values of instellation. For the calculation with instellation \SI{e3}{\watt\per\metre\squared}, the ``Stationary Horizontal'' term is not strongly affected by the imposed background flow until the flow is greater than about \SI{50}{\metre\per\second}. As the jet speed in the corresponding GCM simulation is approximately \SI{38}{\metre\per\second} (as shown in Figure \ref{fig:jet-accn-feedback}), this feedback should not play a large role in the zonal momentum budget. The feedback from the jet is also small in the case with instellation \SI{e2}{\watt\per\metre\squared}, where the Dedalus calculations imply that the jet speed in the relevant GCM simulation is smaller than required to significantly affect the acceleration.

However, Figure \ref{fig:jet-accn-feedback} suggests that this feedback is relevant to the case with instellation $\SI{e4}{\watt\per\metre\squared}$. The calculation implies that the jet speed in this test in the GCM is strong enough to reduce the acceleration to approximately half its strength in the absence of a jet. The jet should still reach equilibrium via the proposed mechanism, but may have a lower maximum speed than predicted. Although this implies that the jet speed could be half that predicted by our estimate (where we neglect this feedback effect), in Section \ref{sec:gcm-simulation-results} we find that our estimate still matches the GCM simulation results reasonably well. This may mean that we have overestimated the effect of this feedback in this section, possibly due to our use of a vertically uniform background flow at the ``jet speed''. In reality, the flow only has this speed at one pressure level, and is significantly less at the peaks of the ``Stationary'' acceleration terms'' (as we explain later, the peak jet speed does not necessarily coincide with either of the maxima of these acceleration terms). In addition, in these Dedalus calculations the imposed jet does not significantly affect the vertical structure of the acceleration terms beyond decreasing their magnitude. This is consistent with \citet{tsai2014three}, where the stationary waves Doppler-shifted by an imposed jet retained the same vertical structure.

In summary, this section shows that for a terrestrial tidally locked planet with instellation \SI{e3}{\watt\per\metre\squared} the feedback of the jet on the stationary waves and the resulting ``Stationary'' acceleration terms is negligible for our simulations with instellation of \SI{e3}{\watt\per\metre\squared} and below. It may be significant for the simulations with instellation higher than this, but we find later that the effect may not be as great as predicted by our idealised calculations in this section, which apply the ``peak'' jet speed at all vertical levels and so may overestimate the effect of this feedback.


\section{Predicting Jet Speed}\label{sec:predicting-jet-speed}

In this section, we estimate the magnitude of the equatorial acceleration for given planetary parameters, and use this to derive the jet speed that will balance this acceleration. In Section \ref{sec:3d-stationary-wave-response}, we predicted the equilibrium jet speed in Equation \ref{eqn:predicted-jet-speed} by estimating the zonal flow that would produce a ``Mean Vertical'' acceleration that would balance the positive ``Stationary Vertical'' acceleration peak shown in Figure \ref{fig:accn-profiles-p}. In this section, we approximate Equation \ref{eqn:predicted-jet-speed} as:

\begin{equation}\label{eqn:predicted-jet-speed-approx}
  \overline{u} \sim - \frac{H}{\overline{w}} \frac{\partial}{\partial p}\left(\overline{u^{*} \omega^{*}}\right).
\end{equation}

This required us to approximate that the numerator (the ``Stationary'' terms) and the denominator (the zonal-mean vertical velocity) of the integrand have a similar vertical profile. This is supported by Figure \ref{fig:accn-profiles-p-nonlin}, where the sum of the ``Stationary'' terms has a similar vertical profile to the forcing $Q$ used in Section \ref{sec:3d-stationary-wave-response}. This means that the vertically varying parts of the numerator and denominator approximately cancel, and the integral in Equation \ref{eqn:predicted-jet-speed} becomes a multiplication by $H$.

To estimate the speed of the jet, we find the magnitude of the velocity perturbations $u^{*},v^{*},w^{*}$ and the zonal-mean vertical velocity $\overline{w}$, for a given forcing $Q$. The behaviour of the system is partly governed by the dominant balance in the thermodynamic equation in Equation \ref{eqn:dedalus-equations}. In this study, the dominant balance is between the $w$ term and the forcing $Q$ \citep{holton2004introduction,vallis2006book}, which is a state known as the ``Weak Temperature Gradient'' (WTG) regime \citep{pierrehumbert2010palette,koll2016temperature,pierrehumbert2018review}. Imposing the WTG approximation and equating these two terms gives

\begin{equation}
    w^{*} \sim \frac{RQ}{N^{2}H}.
\end{equation}

Approximating the derivatives in the continuity equation in Equation \ref{eqn:dedalus-equations} as $\partial / \partial x \sim 1/a$ and $\partial / \partial z \sim 1/H$, gives an estimate of the horizontal velocity perturbation (also known as the ``eddy'' component, the perturbation to the zonal-mean value):

\begin{equation}
    u^{*} \sim \frac{wa}{H} \sim \frac{RaQ}{N^{2}H^{2}}.
\end{equation}

This is the same as the ``advective balance'' in \citet{komacek2016daynighti}, and allows us to estimate the magnitude of the ``Stationary Vertical'' acceleration term in Equation \ref{eqn:predicted-jet-speed-approx}:

\begin{equation}
\begin{aligned}
    \frac{\partial}{\partial p}\left(\overline{u^{*} \omega^{*}}\right) &\sim \frac{\partial}{\partial z}\left(\overline{u^{*} w^{*}}\right) \\
    &\sim u^{*} w^{*} / H \\
    &\sim \frac{aR^{2}Q^{2}}{N^{4}H^{4}}.
\end{aligned}
\end{equation}

This is the same as the magnitude of the ``Stationary Horizontal'' term if it is approximated as $u^{*} u^{*} / a$. This means that the magnitude of the total acceleration in Equation \ref{eqn:predicted-jet-speed-approx} scales in the same way with forcing as the ``Stationary Vertical'' term alone.

We estimate the magnitude of the zonal-mean vertical velocity to predict the jet speed from Equation \ref{eqn:predicted-jet-speed}. We assume that the zonal-mean form of the thermodynamic equation is governed by the same WTG balance:

\begin{equation}
    \overline{w} \sim \frac{R\overline{Q}}{N^{2}H}.
\end{equation}

We also need to estimate the magnitude of the zonal-mean vertical velocity $\overline{w}$. This depends on the zonal mean of the forcing field $Q$. The idealised forcing field used in the calculation in Section \ref{sec:3d-stationary-wave-response} had no zonal mean, but we can estimate the magnitude of an equivalent ``realistic'' field where the night-side forcing is uniform (or zero), as in \citet{perez2013atmospheric}. If we assume that in reality only the day-side forcing contributes to the zonal mean of the forcing (and therefore to the zonal-mean vertical velocity), we can estimate the zonal mean of the forcing to be:

\begin{equation}
    \overline{Q} \sim \frac{1}{2\pi}\int_{-\pi/2}^{\pi/2} Q_{0} dx \sim Q/\pi.
\end{equation}

This gives the following expression for $\overline{w}$:

\begin{equation}
    \overline{w} \sim \frac{RQ}{\pi N^{2}H}.
\end{equation}

Equation \ref{eqn:predicted-jet-speed-approx} then gives the expected jet speed, using the expressions for the magnitude of the acceleration and vertical velocity we have derived:

\begin{equation}
\begin{aligned}
\overline{u} &\sim - \frac{H}{\overline{\omega}} \frac{\partial}{\partial p}\left(\overline{u^{*} \omega^{*}}\right)\\
     &\sim - \frac{H}{\overline{w}} \frac{\partial}{\partial z}\left(\overline{u^{*} w^{*}}\right)\\
    &\sim H  \frac{aR^{2}Q^{2}}{N^{4}H^{4}} / \frac{RQ}{\pi N^{2}H} \\
    &\sim \pi u^{*}.\\
\end{aligned}
\end{equation}

We use the normalisation of the forcing $Q$ from Section \ref{sec:3d-stationary-wave-response} to write this in terms of the instellation:

\begin{equation}\label{eqn:jet-speed-estimate}
    \overline{u} \sim \frac{2.53 \pi ag^{3}\sigma^{1/2}}{RN^{2}p_{0}c_{p}}  F_{0}^{1/2}.
\end{equation}

This corresponds to a maximum jet speed at $z=H$, as by approximating the vertical derivative $\partial\overline{u} / \partial z$ as $\overline{u}/H$ we have assumed that the jet speed increases from 0 at the surface to $\overline{u}$ over the distance $H$. The predicted equatorial jet speed is comparable to the magnitude of the velocity perturbations $\sim u^{*}$ on the equator in a state of nonlinear balance. This arises from the cancellation of the vertical velocity term in the ``Stationary Vertical'' term with the vertical velocity in the denominator of Equation \ref{eqn:predicted-jet-speed}, as well as the cancellation of the vertical derivative in the ``Stationary Vertical'' term with the vertical integral in Equation \ref{eqn:predicted-jet-speed}. It is important to note that $u^{*}$ is not the gravity wave speed $\sqrt{gH}$, as might be expected. Instead, it is the magnitude of the horizontal velocity providing nonlinear zonal momentum balance on the equator. As discussed in Section \ref{sec:idealised-beta-plane-model}, the constant of proportionality in Equation \ref{eqn:jet-speed-estimate} depends on the vertical profile of the heating at the substellar point, so will vary for planets without simple heating profiles (and zero albedo) like those in our GCM simulations. We expect that the overall mechanism and scaling relations will remain the same, unless the heating profile has a scale very different to $H$ or the atmosphere is very thick.

In the next section we will show how this prediction approximately matches the jet speeds predicted by the calculation in Dedalus in Section \ref{sec:3d-stationary-wave-response}, and also matches the GCM results that we will present later.

This prediction of the jet speed can be arrived at even more simply, by interpreting the proposed mechanism as requiring that the magnitude of the acceleration due to ``Stationary Vertical'' transport, $ \frac{\partial}{\partial p}\left(\overline{u}^{*} \overline{\omega}^{*}\right)$, to be equal to the magnitude of the gradient due to the ``Mean Vertical'' interaction between the zonal-mean zonal flow and the zonal-mean vertical velocity, $\overline{\omega} \frac{\partial\overline{u}}{\partial p}$. Setting the ``Stationary Vertical'' and ``Mean Vertical'' terms to be equal, and approximating the derivatives as before gives

\begin{equation}
    \overline{u} \sim \frac{w^{*}}{\overline{w}} u^{*}
\end{equation}

This leads to Equation \ref{eqn:jet-speed-estimate} by the path outlined above. The jet speeds predicted by the calculation in Dedalus and by Equation \ref{eqn:jet-speed-estimate} depend on the value of the Brunt–Väisälä frequency $N_{*}$. These values are therefore only as accurate as the approximation of a constant $N_{*}$, and the accuracy of the value of the chosen $N_{*}$. 

The jet speed is strongly sensitive to the acceleration due to gravity $g$, scaling with $g^{3}$. Each of these factors of $g$ enters through the scale height $H$, which affects the magnitude of the vertical velocity, the horizontal velocity, and determines the vertical scale of the forcing. This cubic dependence on $g$ is misleading, as the Brunt–Väisälä frequency for an (isothermal) atmosphere depends on $g^{2}$, which means that in reality the jet speed only depends linearly on $g$. We have not replaced $N_{*}$ with the exact expression for an isothermal atmosphere in Equation \ref{eqn:jet-speed-estimate} as this is not a good estimate for a realistic terrestrial atmosphere, which is far from isothermal. This again shows the importance of an accurate estimate of the Brunt–Väisälä frequency in estimating the stationary wave strength and the resulting jet speed.

The jet speed depends inversely on $p_{s}$, implying that a very high surface pressure gives a very weak jet. This comes about because we assume that the jet forms at a pressure level comparable to $p_{s}$ on terrestrial planets. For a gaseous planet without a set surface pressure, it would instead be more appropriate to use a pressure level at which the atmosphere is heated strongly by shortwave radiation. In summary, this estimate of jet speed relies on several assumptions about the planet and its atmosphere, and will not apply to planets with different properties. We will investigate the jet speed on gaseous tidally locked planets in a study to follow.

\section{GCM Simulations of Jet Formation}\label{sec:gcm-simulation-results}

\begin{figure*}
\centering 
\subfloat[Equatorial zonal-mean zonal velocity profile of the first 10 days]{%
  \includegraphics[width=\columnwidth]{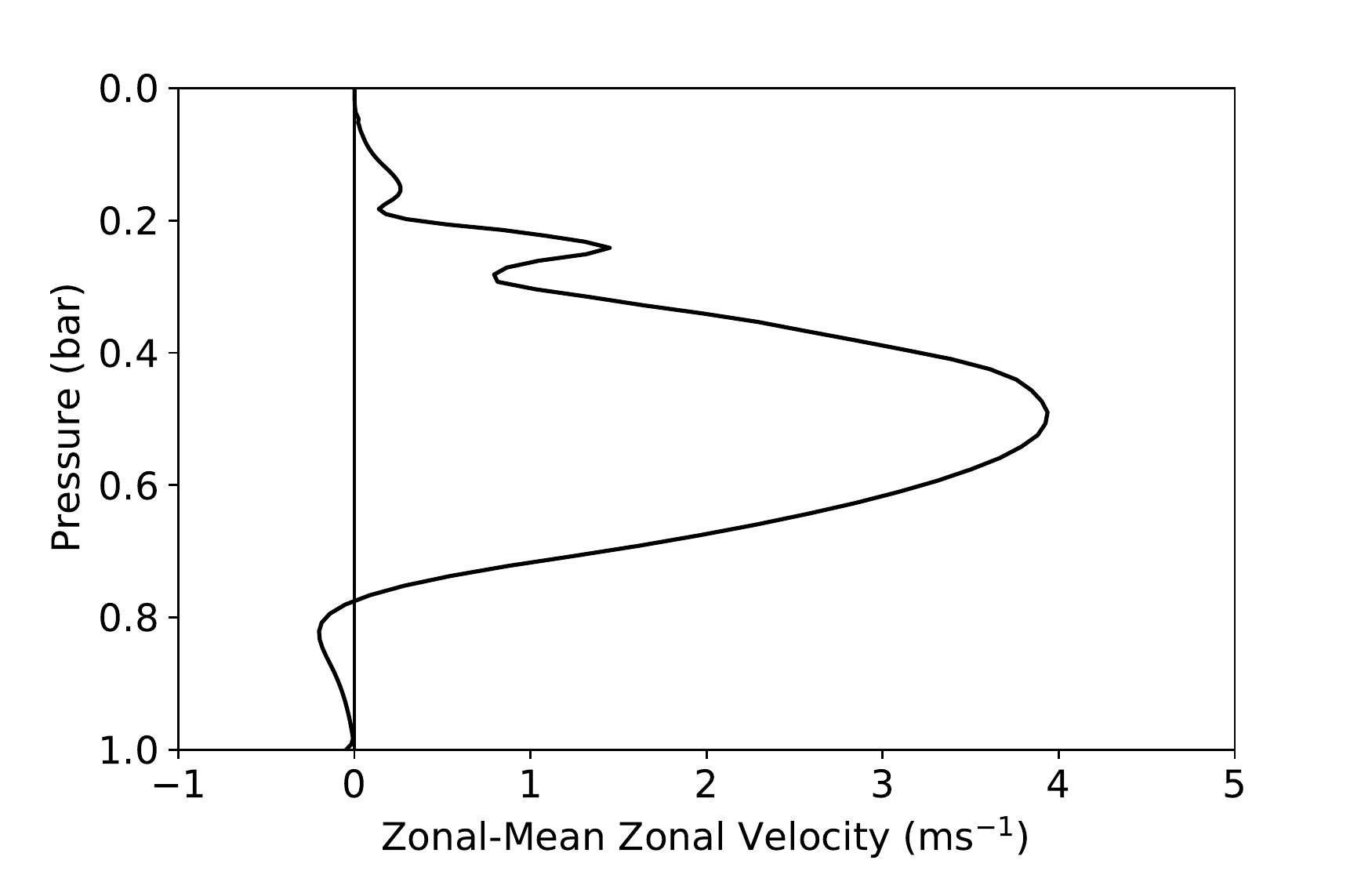}%
  \label{fig:gcm-1000-nonlin-velocity-profile-start}%
}\qquad
\subfloat[Equatorial zonal momentum budget of the first 10 days]{%
  \includegraphics[width=\columnwidth]{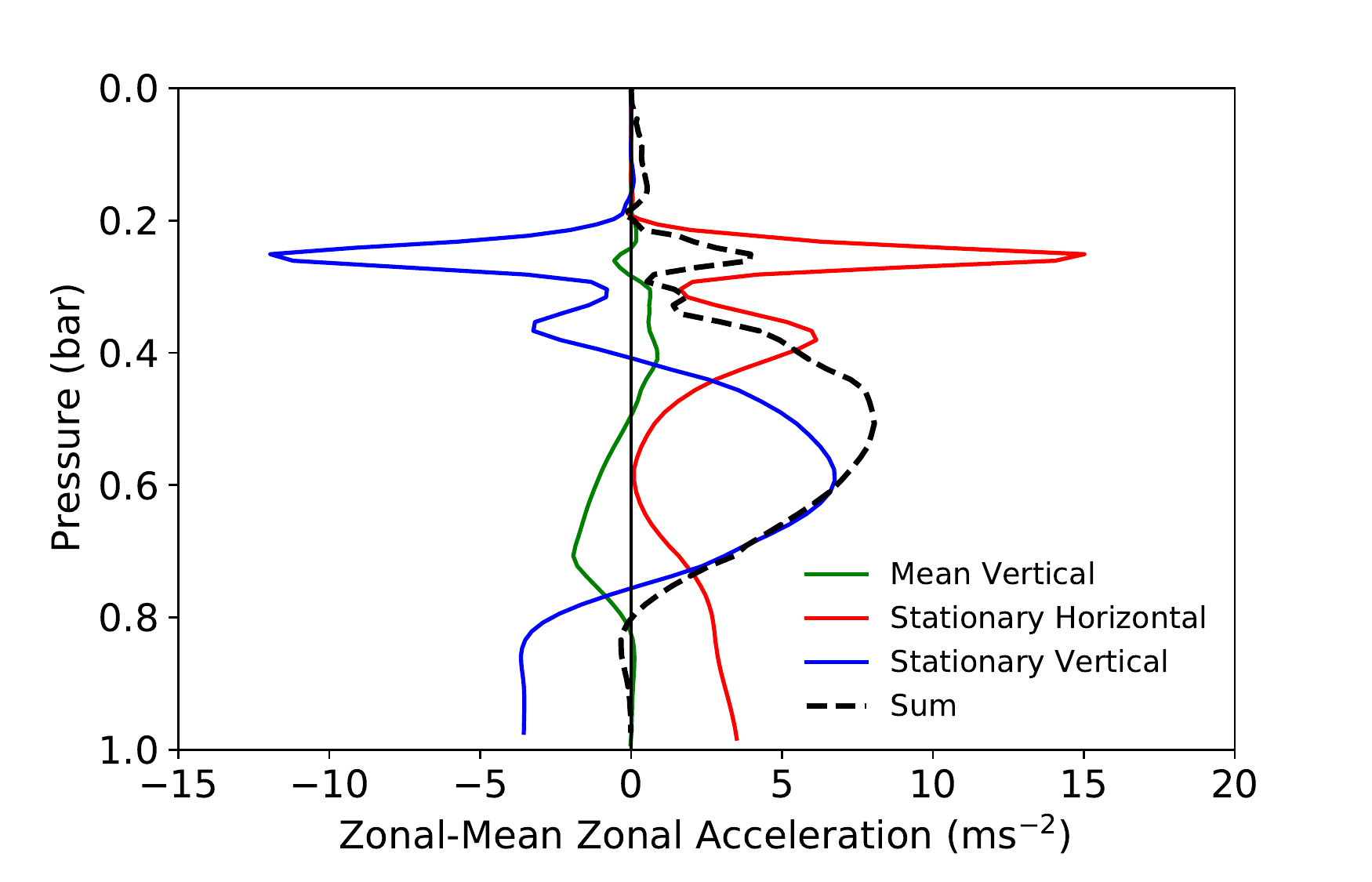}%
  \label{fig:gcm-1000-nonlin-accn-profile-start}%
}
\caption{The jet profile and acceleration terms averaged over the first 10 days of the GCM simulation with instellation \SI{e3}{\watt\per\metre\squared}. The shape of the zonal-mean velocity profile matches the shape of the sum of the ``Stationary'' acceleration terms, supporting our proposed mechanism, where the the initial acceleration is only due to these terms.}\label{fig:gcm-1000-accn-profile-start}
\end{figure*}

In this section, we examine how well the preceding reasoning fares in explaining the zonal momentum budget of GCM simulations. It is to be hoped that at some stage in the future,  the hierarchy of understanding can be completed by comparing the GCM simulations to observations of terrestrial tide-locked planets. We will show that the jet speed in the GCM simulations can be approximately predicted by our previous calculations of the stationary wave response.

\subsection{Numerical simulations}\label{sec:numerical-simulations}

We ran simulations of the atmospheres of terrestrial tidally locked planets in the general circulation model (GCM) ExoFMS \citep{ding2016convection, pierrehumbert2016dynamics}, built on the GFDL FMS (Flexible Modelling System) and the associated cubed-sphere dynamical core \citep{lin2004fv}. This solves the hydrostatic primitive equations on a cubed-sphere grid with a hybrid ``sigma-pressure'' coordinate system in the vertical, and uses a semi-grey radiative transfer scheme and a dry convective adjustment scheme \citep{pierrehumbert2010principles,hammond2018wavemean}. To generate plots, the simulation results are regridded from the cubed-sphere grid to a latitude-longitude grid, and interpolated to a pressure grid in the vertical.

We use 100 vertical levels, to better resolve the vertical profiles of the momentum fluxes shown in Figure \ref{fig:gcm-1000-accn-profile}. The mechanism should still apply to simulations with a lower vertical resolution, and this high vertical resolution would be unnecessary for most other studies. We use a cubed-sphere grid with 48 grid cells on each of the six faces, which corresponds approximately to a spectral resolution of T63. We apply a fourth-order horizontal divergence damping to the velocity fields in the model, with an Earth-like non-dimensional damping coefficient of 0.16 \citep{lin2004fv}.

The planetary atmospheres simulated all have a radius of \SI{6e6}{\metre}, a rotation period of \SI{10}{\day}, acceleration due to gravity of \SI{10}{\metre\per\second\squared}, specific heat capacity \SI{e3}{\joule\per\kilogram\per\kelvin}, molar mass \SI{28}{\gram\per\mol}, and surface pressure \SI{e5}{\pascal}, in order to match the stationary wave calculations in Dedalus in Section \ref{sec:3d-stationary-wave-response}. The simulations have different values of instellation at their substellar point, from \SI{e2}{\watt\per\metre\squared} to \SI{e4}{\watt\per\metre\squared}. This range is comparable to the range of instellation values for the Trappist-1 system \citep{gillon2017seven}, which could provide an observational test of the dependence of the strength of global circulation on instellation. The simulations were spun up for 1000 days to ensure the outgoing longwave radiation and zonal-mean zonal velocity reached equilibrium, then data was recorded every 2 days for 1000 days, to ensure a long enough measurement of the time-mean quantities, with enough time resolution to capture the transient quantities.

\subsection{Equilibrium Momentum Budget}

\begin{figure*}
\centering 
\subfloat[Equilibrium equatorial zonal-mean zonal velocity profile]{%
  \includegraphics[width=\columnwidth]{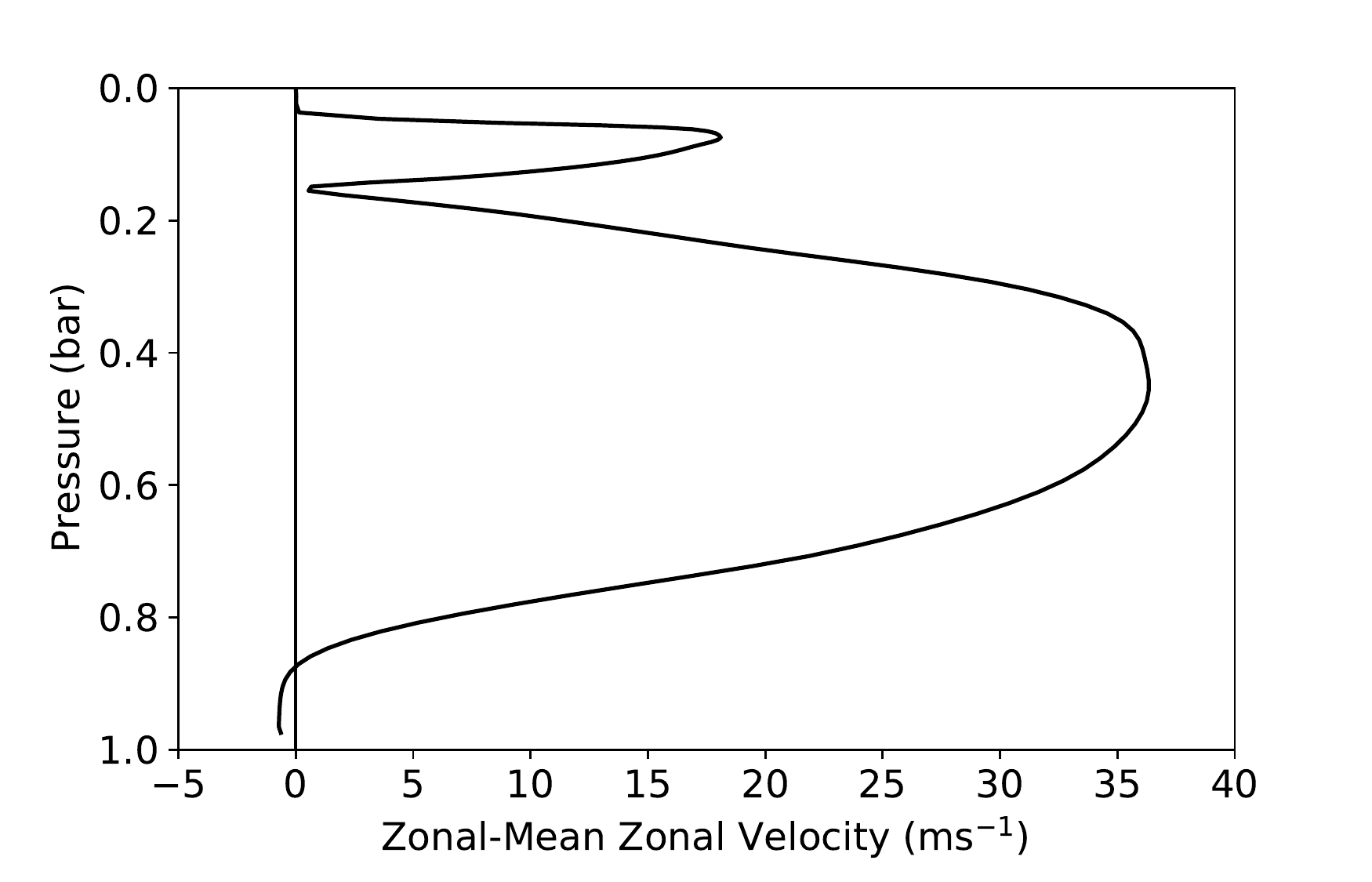}%
  \label{fig:gcm-1000-nonlin-velocity-profile}%
}\qquad
\subfloat[Equilibrium equatorial zonal momentum budget]{%
  \includegraphics[width=\columnwidth]{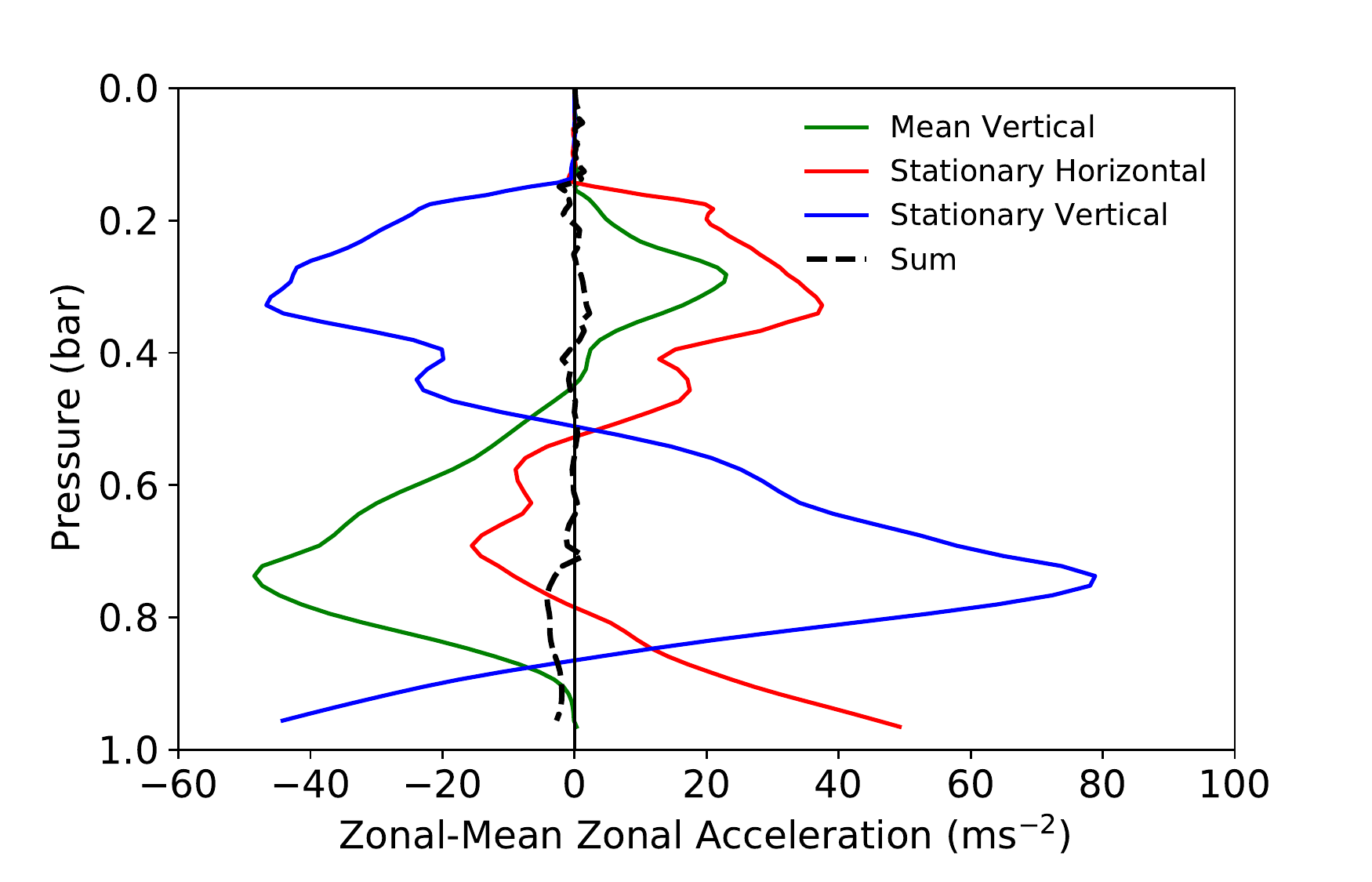}%
  \label{fig:gcm-1000-nonlin-accn-profile}%
}
\caption{The equilibrium jet profile and acceleration terms for the simulation with instellation \SI{e3}{\watt\per\metre\squared} in the GCM. The jet peaks at about $p/p_{s} = 0.4$. The zonal momentum budget is dominated by the balance described in Section \ref{sec:jet-formation-mechanism}, where the ``Stationary'' terms produce a jet that provides a ``Mean Vertical'' acceleration that balances them.}\label{fig:gcm-1000-accn-profile}
\end{figure*}

Figure \ref{fig:gcm-1000-accn-profile-start} shows the state of the model in the first 10 days of the spin-up of the test with instellation \SI{e3}{\watt\per\metre\squared}. The ``Mean Vertical'' term is very small, as the equatorial jet is very weak at this early stage. The shape of the zonal-mean zonal velocity profile in Figure \ref{fig:gcm-1000-nonlin-velocity-profile-start} almost exactly matches the shape of the sum of the two ``Stationary'' terms in Figure \ref{fig:gcm-1000-nonlin-accn-profile-start}. This is a key result that supports our proposed mechanism, where the initial jet is only due to these ``Stationary'' terms. Note that these terms have not yet reached their full strength in equilibrium; this is because the stationary velocity perturbations do not reach their maximum amplitude immediately and take some time to evolve. This means that rather than the ``Stationary'' acceleration terms forming completely, and then the jet forming in response, the jet instead evolves in tandem with the ``Stationary'' terms. The final balance should still be the same as if the jet only evolved to balance the total ``Stationary'' terms after they formed completely, as long as it does not affect these terms strongly (as shown in Section \ref{sec:jet-feedback-on-accn}).

Figure \ref{fig:gcm-1000-accn-profile} shows the state of the simulation in equilibrium. Figure \ref{fig:gcm-1000-nonlin-velocity-profile} plots the zonal-mean zonal velocity at the equator of this test in equilibrium, showing that its maximum is at about \SI{400}{\milli\bar}. In the mechanism we propose, this equatorial jet forms to balance the unbalanced ``Stationary'' acceleration terms. Figure \ref{fig:gcm-1000-nonlin-accn-profile} shows the terms in the zonal-mean momentum budget (Equation \ref{eqn:zonal-mean-acceleration-reduced}) for this simulation. Note that the ``Mean Vertical'' term is zero at the peak of the jet, where the vertical gradient of the jet is zero. The peak of the jet therefore corresponds to the pressure level where the ``Stationary Horizontal'' and ``Stationary Vertical'' terms originally cancel, matching the two-dimensional shallow-water model of \citet{showman2010superrotation} and \citet{showman2011superrotation}. The amplitude of these two-dimenensional models corresponds to the amplitude of the first baroclinic mode with vertical wavelength $2H$, which is the dominant mode excited by the forcing with wavelength $2H$ that we introduced in Section \ref{sec:3d-stationary-wave-response}.

The result that the peak of the jet corresponds to the pressure level where the peaks of the two ``Stationary'' terms always cancel (in Figures \ref{fig:accn-profiles-p} and \ref{fig:gcm-1000-accn-profile}) is notable. If the jet were equilibrated by a linear Rayleigh drag, the peak of the jet would be at the pressure level that had the highest initial acceleration. Instead, it is at a level that has zero zonal-mean zonal acceleration at the initialisation of the model.

In these GCM simulations the ``Mean Vertical'' term does not totally cancel the peak of the ``Stationary Vertical'' term, unlike in the idealised calculations in Section \ref{sec:3d-stationary-wave-response}. The remainder of the momentum balance in the GCM is mostly due to the Rayleigh drag near the surface (not shown explicitly in the plot, but included in the ``Sum'' line). The ``Transient'' terms in Equation \ref{eqn:zonal-mean-acceleration-unsimplified} also contribute to the momentum balance (and are included in the ``Sum'' term), but are small compared to the other terms. In addition, the ``Stationary Horizontal'' term in equilibrium in Figure \ref{fig:gcm-1000-accn-profile} is negative at around \SI{700}{\milli\bar}, and plays a role in equilibrating the jet.  

This does not match the idealised solutions in Section \ref{sec:3d-stationary-wave-response}, where this term is always positive. This difference appears to arise in the GCM in equilibrium at the pressure level where the vertical shear of the zonal-mean zonal jet is strongest; this could be modifying the stationary wave response by a mechanism similar to that shown by \citet{kato1992external}.This process could be similar to the equilibration process suggested by \citet{tsai2014three} that we discussed in Section \ref{sec:jet-feedback-on-accn}, where the jet increases until it decreases the ``Stationary'' terms enough to reach equilibrium.  

The momentum budgets shown in this section have shown the importance of considering the three-dimensional structure of the stationary waves induced by forcing in these atmospheres, and the resulting vertical structure of the zonal acceleration. The next section shows how the equatorial jet speed scales with the magnitude of the instellation in these simulations, and compares it to the speed predicted by our calculations in Section \ref{sec:3d-stationary-wave-response}.

\subsection{Jet Speed Scaling with Forcing}

\begin{figure*}
\centering 
\subfloat[Velocity profiles for different values of instellation]{%
  \includegraphics[width=\columnwidth]{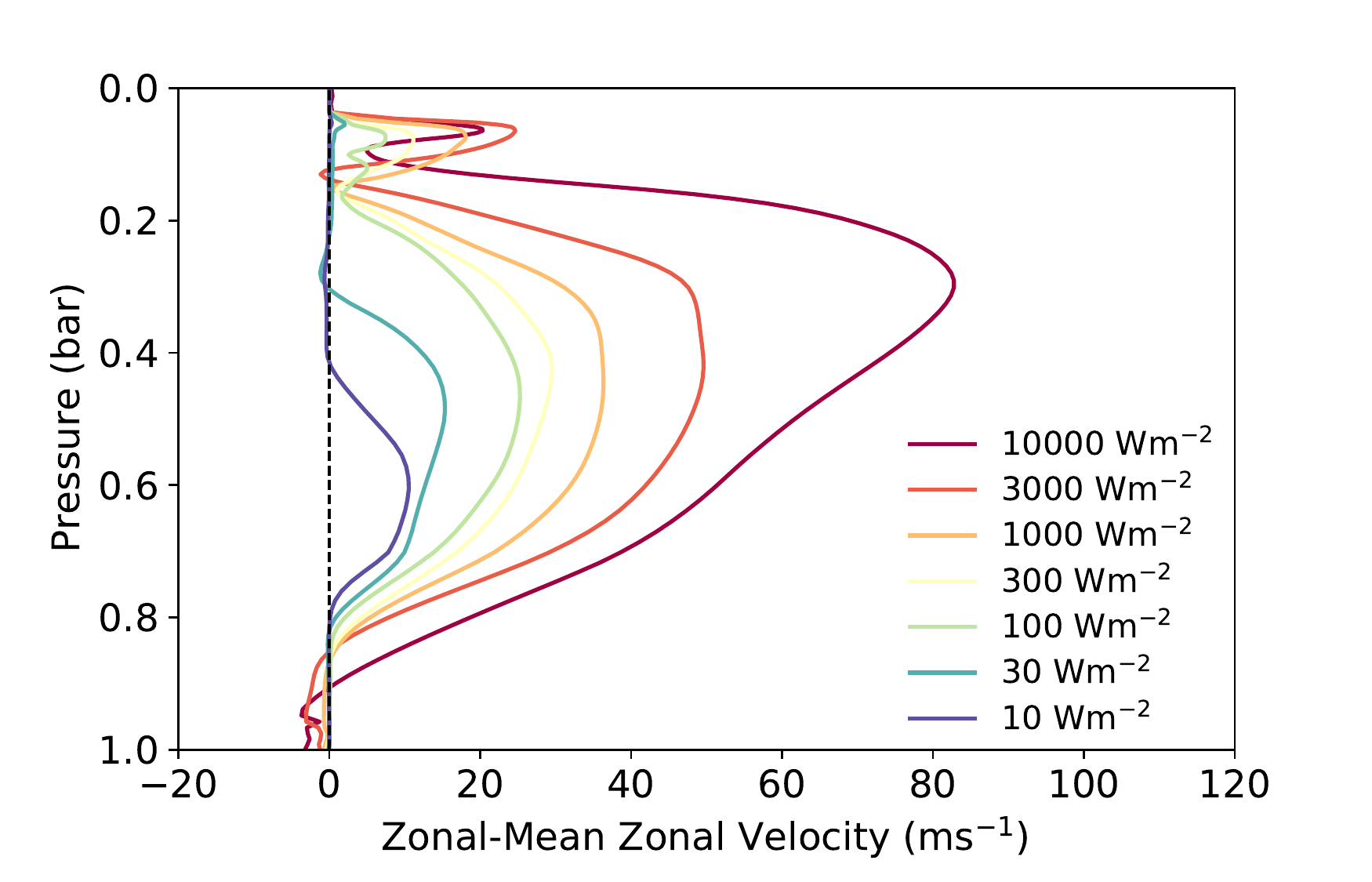}%
  \label{fig:gcm-scaling-nonlin-velocity-profiles}%
}\qquad
\subfloat[Maximum jet speed versus instellation]{%
  \includegraphics[width=\columnwidth]{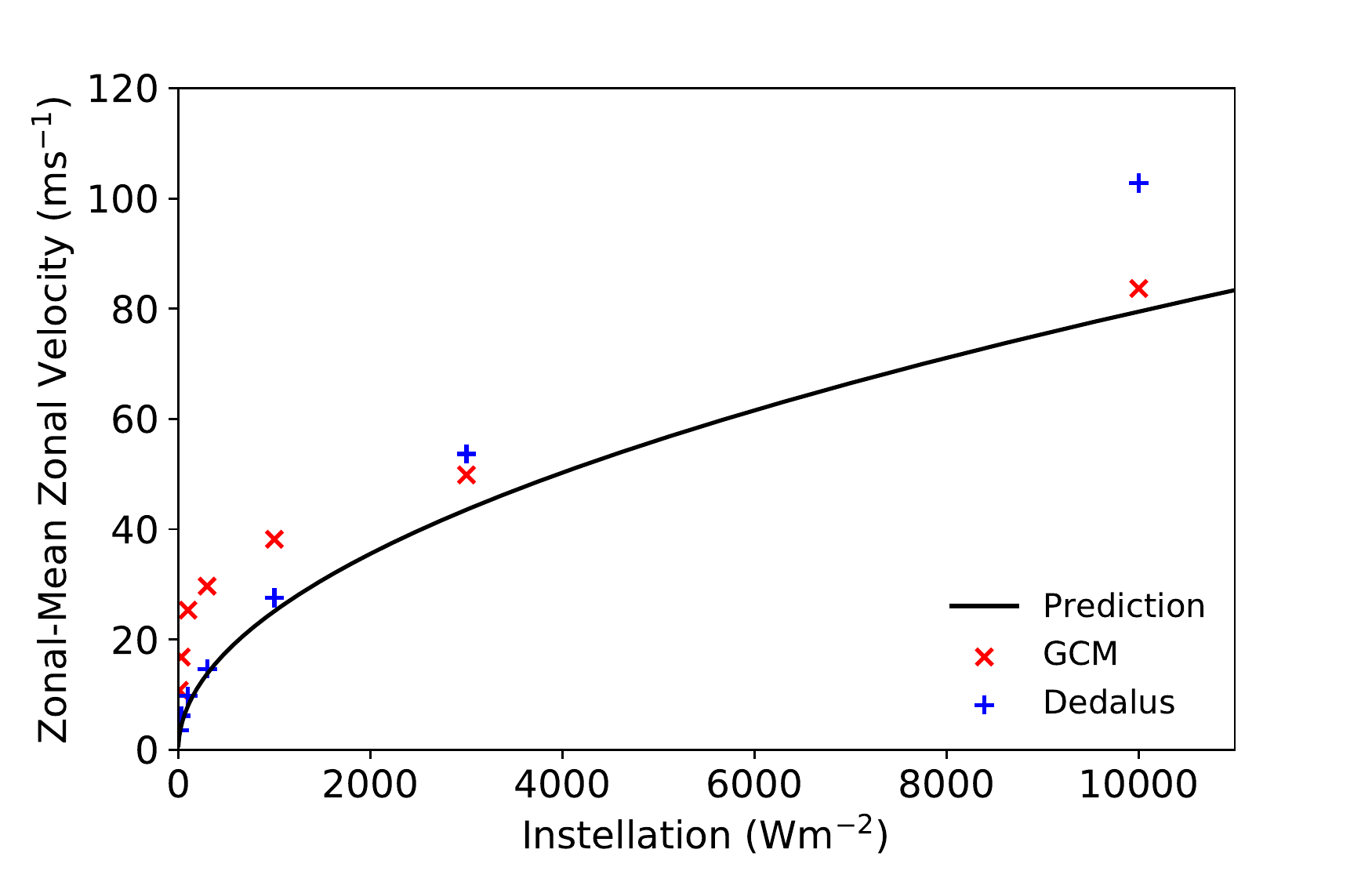}%
  \label{fig:Ubar-scaling-gcm-nonlin}%
}
\caption{The scaling of zonal-mean zonal equatorial velocity versus instellation in our hierarchy of models. The first panel shows how the jet speed depends weakly on instellation in the GCM simulations, scaling approximately with $F_{0}^{1/2}$. The peak of the jet moves to slightly lower pressures at higher values of instellation. This may be due to a larger scale height at higher temperatures, giving a longer vertical scale for the forcing and therefore a longer vertical wavelength for the dominant vertical modes in the stationary waves induced by the forcing. The second panel shows how the maximum jet speed in the GCM simulations scales with instellation, and compares this to the calculation with Dedalus and to the prediction from Equation \ref{eqn:jet-speed-estimate}.}\label{fig:Ubar-scaling-gcm}
\end{figure*}

Figure \ref{fig:Ubar-scaling-gcm} plots the zonal-mean zonal velocity at the equator of each test in the GCM, showing how the jet speed increases weakly with increasing forcing, and how the peak moves to lower pressures. This increase in the height of the jet is likely related to an increase in the atmospheric scale height at higher equilibrium temperatures. 

The atmospheric scale height will increase at higher temperatures, increasing the height of the jet according to the theory in Section \ref{sec:predicting-jet-speed}. However, the idealised scale height $H$ should always correspond to the same pressure $p_{s} e^{-1}$, for surface pressure $p_{s}$. This means that the decrease in the pressure of the core of the jet at higher temperatures seen in Figure \ref{fig:Ubar-scaling-gcm} cannot be explained by an increase in the scale height. It implies that the height of the jet is increasing slightly faster than the scale height $H$ increases with temperature, which cannot be explained by the idealised theory in Section \ref{sec:predicting-jet-speed}. The pressure level of the center of the jet only changes by a factor of 2 over the three orders of magnitude of forcing we model, so we suggest that our approximation that its height is $H$ is fairly accurate. Modelling the exact height of the jet would require a more sophisticated theory, but would be very useful for observable planets where the effect of the jet is measured in phase curves corresponding to particular pressure levels \citep{parmentier2017handbook}.

The jet speed only depends weakly on forcing -- the simulation with the highest instellation has $10^{3}$ times more energy deposited in the atmosphere than the simulation with the lowest instellation, but its jet is only a few times faster. It might be expected from the linear model in \citet{showman2011superrotation} (the case when their forcing is weak) that a forcing $10^{3}$ times stronger would produce velocity perturbations $10^{3}$ times stronger as they are linearly related in that model. This would give an acceleration $10^{6}$ times stronger, giving a jet $10^{6}$ times faster. This is not the case in our GCM simulations, as the nonlinear balance at the equator means that the magnitude of the equatorial acceleration depends more weakly on the forcing. In the mechanism we propose, the equilibrium jet speed also depends inversely on the zonal-mean vertical velocity, which increases with increasing instellation, further reducing the dependence of the jet speed on the instellation.

Figure \ref{fig:Ubar-scaling-gcm-nonlin} compares the results of our hierarchy of models. It shows the maximum jet speed of each test in the GCM, and compares them to the jet speed of the equivalent tests using Dedalus in Section \ref{sec:3d-stationary-wave-response}, as well as the jet speed predicted by Equation \ref{eqn:jet-speed-estimate}. The Dedalus calculations approximately match the magnitude and scaling of the jet speeds in the GCM simulations, suggesting that the mechanism we propose is a good description of the process for jet formation in the GCM. The speed predicted by Equation \ref{eqn:jet-speed-estimate} also approximately matches the speeds in the GCM, suggesting that this is a reasonable estimate of the jet speed on these planets (given the approximations discussed in Section \ref{sec:predicting-jet-speed}).

The jet speed in the GCM scales less strongly than the prediction of Equation \ref{eqn:jet-speed-estimate} at high instellations. This may be due to the feedback discussed in Section \ref{sec:jet-feedback-on-accn}, where the faster jet at higher instellations reduces the magnitude of the ``Stationary Horizontal'' acceleration term that transports momentum towards the equator. A higher instellation and temperature will also increase the relative importance of the nonlinear terms in the thermodynamic equation which we neglected in our derivation of Equation \ref{eqn:jet-speed-estimate}. This will reduce the resulting temperature perturbation and stationary wave strength, reducing the jet speed in the GCM relative to our simple estimate. At the lower end, the jet speed increases more rapidly than predicted by Equation \ref{eqn:jet-speed-estimate}. We could not find a convincing explanation for this, but suggest that it could be due to a linear balance, rather than non-linear balance, controlling the magnitude of the stationary waves. At lower instellation, the stationary waves and the jet are closer to the ground and more affected by the linear Rayleigh drag applied there. This rapid increase could also be due to the onset of the resonance identified by \citet{tsai2014three} that we discussed in Section \ref{sec:jet-feedback-on-accn}.

In summary, the hierarchy of models approximately agree, but the exact jet speed deviates from the prediction at very high and low forcing values. Given the model complexity required in other studies to accurately predict a single jet speed in specific planets \citep{held1980nonlinear,leovy1987zonal,zhu2006maintenance}, we suggest that this approximate match over three orders of magnitude in forcing is a reasonable confirmation of the proposed mechanism.

\section{Discussion}\label{sec:discussion}

The mechanism we propose relies on several approximations and assumptions about which processes are dominant in the zonal momentum budget of the atmosphere, and which are negligible. This section discusses the validity of these approximations and considers other processes which could accelerate or decelerate the equatorial jet.

\subsection{Complications to this mechanism}

The mechanism proposed in this paper is an idealised representation of a complex, time-dependent process. We assume that the ``Stationary'' terms in the zonal momentum equations form to their full strength instantly, then the jet evolves over time to balance them. In reality, the temperature and velocity perturbations of the stationary waves produced by the day-night instellation gradient evolve on a timescale of tens of days in the GCM simulations. This is a similar timescale to the evolution of the jet itself, so the ``Stationary'' acceleration terms and the jet evolve in tandem rather than sequentially. The equilibrium zonal momentum balance in Figure \ref{fig:gcm-1000-accn-profile} still behaves as if the jet evolved to balance the fully-formed ``Stationary'' terms, so this approximation is reasonable. We also approximated that the jet does not strongly affect these ``Stationary'' terms once it has formed; we showed this to be true for the terrestrial planets in this study in Section \ref{sec:jet-feedback-on-accn}, but it may not be true for higher temperature planets, or gaseous planets, which might have stronger jets.

Some of the terms in Equation \ref{eqn:zonal-mean-acceleration-unsimplified} that we neglected could affect the jet speed. For example, we did not consider the effect of the ``Transient'' terms, as they were small for the planetary parameters we used. If the travelling waves were comparable in magnitude to the stationary waves we considered, they could accelerate the equatorial jet \citep{read2018superrotation} or decelerate it \citep{showman2015circulation}. Other processes imposed in GCM simulations may also affect the speed, such as the surface drag. This drag extends to \SI{700}{\milli\bar} in our model; if it were to extend higher or if the jet were to form lower, the drag would affect the zonal momentum budget and the speed of the jet.

We have also neglected the effect of the planetary rotation rate, which may affect the jet speed \citep{showman2015circulation,pierrehumbert2018review}. The rotation rate determines the horizontal scale of the stationary waves, and therefore the magnitude of the gradients that lead to the ``Stationary'' acceleration terms. It will also set the horizontal length scale of the acceleration terms, determining the meridional width of the equatorial jet. The calculations in this study assumed that the horizontal scale of the velocity and temperature perturbations was the planetary radius, which is only true in general for atmospheres with sufficiently low Rossby numbers. 

At higher rotation rates (for example, for terrestrial Earth-sized planets with periods of less than one day), the (equatorial) Rossby radius is significantly smaller than the planetary radius, and the meridional scale of the equatorial waves will therefore be smaller as well. This should produce a stronger gradient in the velocity fields, giving a higher acceleration and a faster jet, as shown in \citet{pierrehumbert2018review}. The calculations in this study therefore only apply to planetary atmospheres with stationary waves on the scale of the entire planet. The rotation rate could be included in the estimate of jet speed given by Equation \ref{eqn:jet-speed-estimate}, by replacing the planetary radius $a$ with an estimate of the actual meridional scale of the stationary waves induced by the forcing. The equatorial Rossby radius would be a possibility for this scale, but would give a much stronger dependence on rotation rate than was seen in \citet{showman2015circulation} and \citet{pierrehumbert2018review}.

\subsection{Gaseous Planets}

The atmospheric circulation of gaseous tidally locked exoplanets such as hot Jupiters is generally more easily observed that that of terrestrial planets. This makes predicting the jet speed of these planets more observationally relevant than for terrestrial planets. In a preliminary investigation, we found that not all of the simplifying assumptions we made in this study apply to gaseous planets. Firstly, there appeared to be a non-negligible feedback from the zonal jet on the ``Stationary'' terms in the zonal-mean momentum budget, as investigated in Section \ref{sec:jet-feedback-on-accn} and demonstrated in \citet{tsai2014three}. This effect appeared strongest where the vertical shear of the jet was strongest, which could be due to an interaction with the vertical external mode \citep{kato1992external}.

Second, we found that the assumption of uniformly upward zonal-mean vertical velocity did not apply to the simulation of a hot Jupiters; as in \citet{mayne2017hotjupiter}, there were regions of zonal-mean downwelling on the equator. This affects the ``Mean Vertical'' term that equilibrates the ``Stationary'' terms in our mechanism, which relies on an upward zonal-mean vertical velocity at all vertical levels at the equator. In addition, hot Jupiters may be strongly affected by magnetic drag, which is often represented by a linear Rayleigh drag \citep{perna2010magnetic}. This would modify the mechanism further (in fact, this could make the mechanism simpler as the scaling of the velocities and the equilibration of the jet would be due to this linear drag only). 

We will investigate these issues in a study to follow, and are aiming to produce a simple estimate of the magnitude and scaling of the equatorial jet speed on hot Jupiters. This would provide a basis for estimates of observable quantities, such as the predictions of hot-spot shift and day-night contrast in \citet{zhang2017dynamics}.

\section{Conclusion}\label{sec:conclusion}

This study aimed to explain the formation of the equatorial jet on terrestrial tidally locked planets, and to predict its speed. We proposed that the jet forms by a mechanism in which the day-night forcing induces stationary waves which accelerate a jet, which then interacts with the zonal-mean vertical velocity to produce a deceleration that balances the acceleration from stationary waves in equilibrium. We derived the structure of the zonal acceleration by calculating the three-dimensional stationary wave response to the forcing on a tidally locked planet using the Dedalus software. This calculation allowed us to predict the equatorial jet speed for given planetary parameters.

This mechanism was verified by GCM simulations in the ExoFMS model, where the dominant zonal-mean momentum balance was the same as in the proposed mechanism. We ran a suite of simulations with different values of instellation, and showed that the zonal momentum balance and resulting jet speed was approximately the same as predicted by the idealised calculations using Dedalus. With this confirmation of the proposed mechanism, we derived a simplified expression for the jet speed using the WTG approximation:

\begin{equation}
    \overline{u} \sim \frac{2.53 \pi ag^{3}\sigma^{1/2}}{RN^{2}p_{0}c_{p}}  F_{0}^{1/2}.
\end{equation}

The exact jet speed predicted by this expression depends on parameters such as the Brunt–Väisälä frequency $N_{*}$, which is difficult to estimate accurately in general. The constant of proportionality also depends on the vertical heating profile at the substellar point, and the planetary albedo. However, we expect that the relation $\overline{u} \sim  F_{0}^{1/2}$ should hold as long as the mechanism we propose in this study holds. 

We discussed why this mechanism may not apply to gaseous tidally locked planets such as hot Jupiters, and suggested how it could be modified to describe their behaviour. The feedback of the jet on the ``Stationary'' acceleration terms may need to be taken into account for other types of planet, particularly those with high instellation. The mechanism could be modified in other ways to describe other types of tidally locked planetary atmospheres, such as those with cloud layers or significant moisture. These may have different vertical heating profiles and different thermodynamic balances to those assumed in this study of dry, cloud-free atmospheres.

We conclude that this mechanism describes the formation of the equatorial superrotating jet on terrestrial tidally locked planets. It provides a simple prediction of the approximate jet speed, and explains why the jet speed only depends weakly on instellation. We hope that the prediction for jet speeds will be useful for interpreting observations and directing modelling, and will investigate the equivalent mechanism for the formation of equatorial jets on gaseous tidally locked planets in a study to follow.

\acknowledgments

M.H. acknowledges support from the Lindemann Fellowship and from an S.T.F.C. studentship. R.T.P. and S.-M.T. acknowledge support from the European Research Council Advanced Grant ``Exocondense'' (\#740963). We thank D. Abbot and N. Lewis for helpful discussions. We greatly thank the HPC support staff in the Department of Physics at the University of Oxford.

\newpage

\bibliography{main.bib}

\begin{thebibliography}{}
\expandafter\ifx\csname natexlab\endcsname\relax\def\natexlab#1{#1}\fi
\providecommand{\url}[1]{\href{#1}{#1}}
\providecommand{\dodoi}[1]{doi:~\href{http://doi.org/#1}{\nolinkurl{#1}}}
\providecommand{\doeprint}[1]{\href{http://ascl.net/#1}{\nolinkurl{http://ascl.net/#1}}}
\providecommand{\doarXiv}[1]{\href{https://arxiv.org/abs/#1}{\nolinkurl{https://arxiv.org/abs/#1}}}

\bibitem[{Arcangeli {et~al.}(2019)Arcangeli, D{\'e}sert, Parmentier, Stevenson,
  Bean, Line, Kreidberg, Fortney, \& Showman}]{arcangeli2019climate}
Arcangeli, J., D{\'e}sert, J.-M., Parmentier, V., {et~al.} 2019, Astronomy \&
  Astrophysics, 625, A136

\bibitem[{Boutle {et~al.}(2017)Boutle, Mayne, Drummond, Manners, Goyal,
  Hugo~Lambert, Acreman, \& Earnshaw}]{boutle2017proxima}
Boutle, I.~A., Mayne, N.~J., Drummond, B., {et~al.} 2017, Astronomy {\&}
  Astrophysics, 601, A120

\bibitem[{Brogi {et~al.}(2016)Brogi, De~Kok, Albrecht, Snellen, Birkby, \&
  Schwarz}]{brogi2016rotation}
Brogi, M., De~Kok, R., Albrecht, S., {et~al.} 2016, The Astrophysical Journal,
  817, 106

\bibitem[{Burns {et~al.}(2016)Burns, Vasil, Oishi, Lecoanet, \&
  Brown}]{burns2016dedalus}
Burns, K.~J., Vasil, G.~M., Oishi, J.~S., Lecoanet, D., \& Brown, B. 2016,
  Astrophysics Source Code Library

\bibitem[{Burns {et~al.}(2020)Burns, Vasil, Oishi, Lecoanet, \&
  Brown}]{burns2020dedalus}
Burns, K.~J., Vasil, G.~M., Oishi, J.~S., Lecoanet, D., \& Brown, B.~P. 2020,
  Physical Review Research, 2, 023068

\bibitem[{Carone {et~al.}(2015)Carone, Keppens, \& Decin}]{carone2015regimes}
Carone, L., Keppens, R., \& Decin, L. 2015, Monthly Notices of the Royal
  Astronomical Society, 453, 2412

\bibitem[{Cho {et~al.}(2015)Cho, Polichtchouk, \&
  Thrastarson}]{cho2015sensitivity}
Cho, J.-K., Polichtchouk, I., \& Thrastarson, H.~T. 2015, Monthly Notices of
  the Royal Astronomical Society, 454, 3423

\bibitem[{Debras {et~al.}(2020)Debras, Mayne, Baraffe, Jaupart, Mourier, Laibe,
  Goffrey, \& Thuburn}]{debras2020acceleration}
Debras, F., Mayne, N., Baraffe, I., {et~al.} 2020, Astronomy \& Astrophysics,
  633, A2

\bibitem[{Demory {et~al.}(2016)Demory, Gillon, de~Wit, Madhusudhan, Bolmont,
  Heng, Kataria, Lewis, Hu, Krick, Stamenkovi{\'c}, Benneke, Kane, \&
  Queloz}]{demory201655cnce}
Demory, B.-O., Gillon, M., de~Wit, J., {et~al.} 2016, Nature, 532, 207

\bibitem[{Ding \& Pierrehumbert(2016)}]{ding2016convection}
Ding, F., \& Pierrehumbert, R.~T. 2016, The Astrophysical Journal, 822, 24

\bibitem[{Drummond {et~al.}(2018)Drummond, Mayne, Manners, Carter, Boutle,
  Baraffe, H{\'e}brard, Tremblin, Sing, Amundsen,
  {et~al.}}]{drummond2018observable}
Drummond, B., Mayne, N., Manners, J., {et~al.} 2018, The Astrophysical Journal
  Letters, 855, L31

\bibitem[{Fels \& Lindzen(1974)}]{fels1974interaction}
Fels, S.~B., \& Lindzen, R.~S. 1974, Geophysical and Astrophysical Fluid
  Dynamics, 6, 149

\bibitem[{Flowers {et~al.}(2019)Flowers, Brogi, Rauscher, Kempton, \&
  Chiavassa}]{flowers2019high}
Flowers, E., Brogi, M., Rauscher, E., Kempton, E. M.-R., \& Chiavassa, A. 2019,
  The Astronomical Journal, 157, 209

\bibitem[{Gill \& Philips(1986)}]{gill1986nonlinear}
Gill, A., \& Philips, P. 1986, Quarterly Journal of the Royal Meteorological
  Society, 112, 69

\bibitem[{Gill(1980)}]{gill1980solutions}
Gill, A.~E. 1980, Quarterly Journal of the Royal Meteorological Society, 106,
  447

\bibitem[{Gillon {et~al.}(2017)Gillon, Triaud, Demory, Jehin, Agol, Deck,
  Lederer, De~Wit, Burdanov, Ingalls, {et~al.}}]{gillon2017seven}
Gillon, M., Triaud, A.~H., Demory, B.-O., {et~al.} 2017, Nature, 542, 456

\bibitem[{Hammond \& Pierrehumbert(2017)}]{hammond2017climate}
Hammond, M., \& Pierrehumbert, R.~T. 2017, The Astrophysical Journal, 849, 152

\bibitem[{Hammond \& Pierrehumbert(2018)}]{hammond2018wavemean}
---. 2018, The Astrophysical Journal, 869, 65

\bibitem[{Held \& Hou(1980)}]{held1980nonlinear}
Held, I.~M., \& Hou, A.~Y. 1980, Journal of the Atmospheric Sciences, 37, 515

\bibitem[{Heng \& Workman(2014)}]{heng2014analytical}
Heng, K., \& Workman, J. 2014, Astrophysical Journal, Supplement Series, 213,
  27

\bibitem[{Hide(1969)}]{hide1969dynamics}
Hide, R. 1969, Journal of the Atmospheric Sciences, 26, 841

\bibitem[{Holton(2004)}]{holton2004introduction}
Holton, J.~R. 2004, An introduction to dynamic meteorology (Elsevier Academic
  Press,)

\bibitem[{Jablonowski \& Williamson(2011)}]{jablonowski2011pros}
Jablonowski, C., \& Williamson, D.~L. 2011, in Numerical techniques for global
  atmospheric models (Springer), 381--493

\bibitem[{Joshi {et~al.}(1997)Joshi, Haberle, \& Reynolds}]{joshi1997tidally}
Joshi, M., Haberle, R., \& Reynolds, R. 1997, Icarus, 129, 450

\bibitem[{Kataria {et~al.}(2014)Kataria, Showman, Fortney, Marley, \&
  Freedman}]{kataria2014atmospheric}
Kataria, T., Showman, A.~P., Fortney, J.~J., Marley, M.~S., \& Freedman, R.~S.
  2014, The Astrophysical Journal, 785, 92

\bibitem[{Kato \& Matsuda(1992)}]{kato1992external}
Kato, T., \& Matsuda, Y. 1992, Journal of the Meteorological Society of Japan.
  Ser. II, 70, 1057

\bibitem[{Koll \& Komacek(2018)}]{koll2018heatengine}
Koll, D.~D., \& Komacek, T.~D. 2018, The Astrophysical Journal, 853, 133

\bibitem[{Koll \& Abbot(2015)}]{koll2015phasecurves}
Koll, D. D.~B., \& Abbot, D.~S. 2015, The Astrophysical Journal, 802, 21

\bibitem[{Koll \& Abbot(2016)}]{koll2016temperature}
---. 2016, The Astrophysical Journal, 825, 99

\bibitem[{Komacek \& Showman(2016)}]{komacek2016daynighti}
Komacek, T.~D., \& Showman, A.~P. 2016, The Astrophysical Journal, 821, 16

\bibitem[{Komacek {et~al.}(2017)Komacek, Showman, \&
  Tan}]{komacek2017daynightii}
Komacek, T.~D., Showman, A.~P., \& Tan, X. 2017, The Astrophysical Journal,
  835, 198

\bibitem[{Kreidberg {et~al.}(2019)Kreidberg, Koll, Morley, Hu, Schaefer,
  Deming, Stevenson, Dittmann, Vanderburg, Berardo, Guo, Stassun, Crossfield,
  Charbonneau, Latham, Loeb, Ricker, Seager, \& Vanderspek}]{kreidberg2019lhs}
Kreidberg, L., Koll, D. D.~B., Morley, C., {et~al.} 2019, Nature

\bibitem[{Leovy(1987)}]{leovy1987zonal}
Leovy, C.~B. 1987, Icarus, 69, 193

\bibitem[{Lin(2004)}]{lin2004fv}
Lin, S.-J. 2004, Monthly Weather Review, 132, 2293

\bibitem[{Louden \& Wheatley(2015)}]{louden2015spatially}
Louden, T., \& Wheatley, P.~J. 2015, The Astrophysical Journal Letters, 814,
  L24

\bibitem[{Lutsko(2018)}]{lutsko2018response}
Lutsko, N.~J. 2018, Journal of the Atmospheric Sciences, 75, 3

\bibitem[{Majeau {et~al.}(2012)Majeau, Agol, \& Cowan}]{majeau20122dmap}
Majeau, C., Agol, E., \& Cowan, N.~B. 2012, The Astrophysical Journal Letters,
  747, L20

\bibitem[{Matsuno(1966)}]{matsuno1966quasi}
Matsuno, T. 1966, Journal of the Meteorological Society of Japan Ser II, 44, 25

\bibitem[{Mayne {et~al.}(2019)Mayne, Drummond, Debras, Jaupart, Manners,
  Boutle, Baraffe, \& Kohary}]{mayne2019limits}
Mayne, N.~J., Drummond, B., Debras, F., {et~al.} 2019, The Astrophysical
  Journal, 871, 56

\bibitem[{Mayne {et~al.}(2014)Mayne, Baraffe, Acreman, Smith, Browning,
  Amundsen, Wood, Thuburn, \& Jackson}]{mayne2014unified}
Mayne, N.~J., Baraffe, I., Acreman, D.~M., {et~al.} 2014, Astronomy \&
  Astrophysics, 561, A1

\bibitem[{Mayne {et~al.}(2017)Mayne, Debras, Baraffe, Thuburn, Amundsen,
  Acreman, Smith, Browning, Manners, \& Wood}]{mayne2017hotjupiter}
Mayne, N.~J., Debras, F., Baraffe, I., {et~al.} 2017, Astronomy {\&}
  Astrophysics, 604, A79

\bibitem[{Mendon{\c{c}}a(2020)}]{mendoncca2020angular}
Mendon{\c{c}}a, J.~M. 2020, Monthly Notices of the Royal Astronomical Society,
  491, 1456

\bibitem[{Mendon{\c{c}}a {et~al.}(2018)Mendon{\c{c}}a, Malik, Demory, \&
  Heng}]{mendoncca2018revisiting}
Mendon{\c{c}}a, J.~M., Malik, M., Demory, B.-O., \& Heng, K. 2018, The
  Astronomical Journal, 155, 150

\bibitem[{Norton(2006)}]{norton2006tropical}
Norton, W. 2006, Journal of the atmospheric sciences, 63, 1420

\bibitem[{Parmentier \& Crossfield(2017)}]{parmentier2017handbook}
Parmentier, V., \& Crossfield, I. J.~M. 2017, Handbook of Exoplanets, 564, 116

\bibitem[{Perez-Becker \& Showman(2013)}]{perez2013atmospheric}
Perez-Becker, D., \& Showman, A.~P. 2013, The Astrophysical Journal, 776, 134

\bibitem[{Perna {et~al.}(2010)Perna, Menou, \& Rauscher}]{perna2010magnetic}
Perna, R., Menou, K., \& Rauscher, E. 2010, The Astrophysical Journal, 719,
  1421

\bibitem[{Pierrehumbert(2010{\natexlab{a}})}]{pierrehumbert2010palette}
Pierrehumbert, R.~T. 2010{\natexlab{a}}, The Astrophysical Journal Letters,
  726, L8

\bibitem[{Pierrehumbert(2010{\natexlab{b}})}]{pierrehumbert2010principles}
---. 2010{\natexlab{b}}, {Principles of Planetary Climate} (Cambridge
  University Press)

\bibitem[{Pierrehumbert \& Ding(2016)}]{pierrehumbert2016dynamics}
Pierrehumbert, R.~T., \& Ding, F. 2016, Proceedings of the Royal Society A:
  Mathematical, Physical and Engineering Sciences, 472, 20160107

\bibitem[{Pierrehumbert \& Hammond(2019)}]{pierrehumbert2018review}
Pierrehumbert, R.~T., \& Hammond, M. 2019, Annual Review of Fluid Mechanics,
  51, 275

\bibitem[{Read \& Lebonnois(2018)}]{read2018superrotation}
Read, P.~L., \& Lebonnois, S. 2018, Annual Review of Earth and Planetary
  Sciences, 46, 175

\bibitem[{Shell \& Held(2004)}]{shell2004superrotation}
Shell, K.~M., \& Held, I.~M. 2004, Journal of Atmospheric Sciences, 61, 2928

\bibitem[{Showman {et~al.}(2015)Showman, Lewis, \&
  Fortney}]{showman2015circulation}
Showman, A.~P., Lewis, N.~K., \& Fortney, J.~J. 2015, Astrophysical Journal,
  801, 95

\bibitem[{Showman \& Polvani(2010)}]{showman2010superrotation}
Showman, A.~P., \& Polvani, L.~M. 2010, Geophysical Research Letters, 37

\bibitem[{Showman \& Polvani(2011)}]{showman2011superrotation}
---. 2011, The Astrophysical Journal, 738, 71

\bibitem[{Stevenson {et~al.}(2014)Stevenson, D{\'e}sert, Line, Bean, Fortney,
  Showman, Kataria, Kreidberg, McCullough, Henry,
  {et~al.}}]{stevenson2014thermal}
Stevenson, K.~B., D{\'e}sert, J.-M., Line, M.~R., {et~al.} 2014, Science, 346,
  838

\bibitem[{Tsai {et~al.}(2014)Tsai, Dobbs-Dixon, \& Gu}]{tsai2014three}
Tsai, S.~M., Dobbs-Dixon, I., \& Gu, P.~G. 2014, Astrophysical Journal, 793,
  141

\bibitem[{Vallis(2006)}]{vallis2006book}
Vallis, G.~K. 2006, Atmospheric and Oceanic Fluid Dynamics, 770

\bibitem[{Wu {et~al.}(2000)Wu, Sarachik, \& Battisti}]{wu2000vertical}
Wu, Z., Sarachik, E.~S., \& Battisti, D.~S. 2000, Journal of the Atmospheric
  Sciences, 57, 2169

\bibitem[{Zhang \& Showman(2017)}]{zhang2017dynamics}
Zhang, X., \& Showman, A.~P. 2017, The Astrophysical Journal, 836, 73

\bibitem[{Zhu(2006)}]{zhu2006maintenance}
Zhu, X. 2006, Planetary and Space Science, 54, 761

\end{thebibliography}

\end{document}